\documentclass[%
 reprint,
 amsmath,amssymb,
 aps,
 superscriptaddress,
]{revtex4-1}

\usepackage{graphicx}
\usepackage{dcolumn}
\usepackage{bm}
\usepackage{hyperref}
\usepackage{xcolor}
\usepackage{amsmath,amsfonts}
\usepackage{comment}
\hypersetup{
    colorlinks=true,
    citecolor = blue,
    linkcolor=blue,
    filecolor=magenta,      
    urlcolor=blue,
    }

\newcommand{\rob}[1]{{\color{black} #1}}
\newcommand{\robnew}[1]{{\color{black} #1}} 
\newcommand{\robn}[1]{{\color{black} #1}}
\newcommand{\mike}[1]{{\color{black} #1}}
\newcommand{\edit}[1]{{\color{black} #1}} 

\begin{document}

\title{Experimentally bounding deviations from quantum theory for a photonic three-level system using theory-agnostic tomography}

\author{Michael J. Grabowecky}
 \address{%
 Institute for Quantum Computing and Department of Physics \& Astronomy,
 University of Waterloo, Waterloo, Ontario, N2L 3G1, Canada}
 \email{mgrabowecky@uwaterloo.ca}

\author{Christopher A. J. Pollack}
 \address{%
 Institute for Quantum Computing and Department of Physics \& Astronomy,
 University of Waterloo, Waterloo, Ontario, N2L 3G1, Canada}
 \address{%
 Department of Mathematics,
 University of Toronto, Toronto, Ontario, M5S 1A1, Canada}
\author{Andrew R. Cameron}
 \address{%
 Institute for Quantum Computing and Department of Physics \& Astronomy,
 University of Waterloo, Waterloo, Ontario, N2L 3G1, Canada}
\author{\\ Robert W. Spekkens}

\address{Perimeter Institute for Theoretical Physics, 31 Caroline Street North, Waterloo, Ontario, N2L 2Y5}
\author{Kevin J. Resch}%

 \address{%
 Institute for Quantum Computing and Department of Physics \& Astronomy,
 University of Waterloo, Waterloo, Ontario, N2L 3G1, Canada}


\begin{abstract}

If one seeks to test quantum theory \rob{against many alternatives in a landscape of possible physical theories,} 
 then it is crucial to be able to analyze experimental data in a theory-agnostic way. This can be achieved using the framework of Generalized Probabilistic Theories (GPTs). 
Here, \rob{we implement GPT tomography on}
  a three-level system \rob{corresponding to a single photon shared among three modes.}
This scheme achieves a GPT characterization of each of the preparations and measurements implemented in the experiment without requiring any prior characterization of either. \rob{Assuming that the sets of realized preparations and measurements are tomographically complete,} our analysis identifies the most likely dimension of the GPT vector space describing the three-level system to be \rob{nine, in agreement} with the value predicted by quantum theory. \rob{Relative to this dimension,} we infer the scope of GPTs that are consistent with our experimental data \rob{by identifying polytopes that provide inner and outer bounds for}
  the state and effect spaces of the true GPT.
   \rob{From these, we are able to determine}
    quantitative bounds on possible deviations from quantum theory.  In particular, we bound the degree to which the no-restriction hypothesis might be violated for our three-level system. 
\end{abstract}

\pacs{Valid PACS appear here}
\maketitle

\section{Introduction}
Despite the fact that quantum theory has thus far been very successful in describing nature, it may one day be superseded by a novel post-quantum theory. In recent years, there have been many theoretical milestones in identifying plausible cadidates for such a theory \cite{Masanes_2011,Sainz2018,Chiribella2011}. Some of these theories can even be explained within the quantum formalism itself, such as models that exhibit intrinsic decoherence \cite{1986,Percival,Stamp,PhysRevA.44.5401}. However, most theories require a rejection of some or all of the quantum framework. Some important examples include Almost Quantum Theory \cite{Sainz2018,PhysRevLett.120.200402}, theories involving higher order interference \cite{doi:10.1142/S021773239400294X,Sinha418,Park_2012,Kauten_2017}, and theories involving quaternions \cite{PhysRevLett.42.683,ADLER1994195}. 

In contrast to the \rob{abundance}
 of theoretical proposals, there \rob{have been few experiments that aim to} 
  test \rob{quantum theory simultaneously against many alternative physical theories.}
 \rob{Doing so} is difficult as one cannot \rob{a priori} assume the validity of any \rob{one particular theory}.
 For this reason, a theory-neutral framework must be adopted to analyze the experimental data. Such a framework is provided by the formalism of {\em generalized probabilistic theories} (GPTs).
 

 The GPT formalism provides a description of a physical system from an operational approach \cite{Hardy2011,Chiribella2010,Barrett2007,Hardy2001,Scandolo2019,Weilenmann2018,d'ariano_2010,Short_2010,Masanes_2011,Janotta_2014} and is used frequently in the field of quantum foundations. It is operational because it describes the theory based solely on what it predicts for the probabilities for each outcome of a measurement in an experiment. 
 \rob{This} framework requires the use of two weak assumptions, both of which are adopted in standard quantum theory. The first is that each choice of preparation is made independently from each choice of measurement. The other assumption is that repeated runs of the same experiment yields an independent and identically distributed (i.i.d.) source of data. Under the validity of these two assumptions, the GPT framework is completely general for modelling experimental data, allowing one to avoid any implicit bias towards quantum theory.
 
 Recent work placed bounds on possible deviations from quantum theory within a landscape of alternative theories for a two-level system using the GPT framework~\cite{Mazurek2017}. In that work, the two-level system was the polarization degree of freedom of a single photon. \robnew{A large number of repetitions of an experiment on photon polarization were conducted, across which the preparation procedure and the measurement procedure was varied (from among a large set of possibilities for each).} 
 The authors developed a \robnew{scheme for achieving a GPT characterization of these preparations and measurements, that is, a {\em }GPT tomography scheme. This scheme requires no prior characterization of the preparations or measurements.  Rather, these are simultaneously characterized based on their interplay.}
 We will refer to this scheme as \textit{\robnew{bootstrap} GPT tomography}.
 \robnew{Bootstrap GPT tomography involves {\em two} key innovations relative to standard quantum tomography~\cite{james, Lundeen2009}. 
 First, standard quantum tomography schemes are only valid under the assumption of the correctness of quantum theory. They do not return {\em GPT} characterizations of the states and measurements, but only {\em quantum} characterizations thereof.  In this sense, they fail to be {\em theory-agnostic}.  Second, standard schemes achieve a characterization of the states under the assumption of a prior characterization of the measurements~\cite{james}, or they achieve a characterization of the measurements 
 under the assumption of a prior characterization of the states~\cite{Lundeen2009}.  They do not achieve a {\em bootstrap} characterization of these, that is, a simultaneous characterization of both states and measurements based on their interplay.
 }
 
 In the present work, we apply \robnew{bootstrap} GPT tomography to a photonic three-level system. The constraints describing the geometry of the quantum state and effect spaces of three-level systems are more complex when compared with their two-level counterparts \cite{Weilenmann2020analysingcausal}. Due to this complexity, there is a greater opportunity for an experiment to reveal some surprising discrepancies between nature and the predictions of quantum theory \footnote{for instance, Ref. \cite{Rudolph2006} proposes a modification to quantum theory that does not impact qubits, but that alters the predictions for qutrits}. 
 
 Our experiment is conducted using heralded single photons prepared using spontaneous parametric down conversion (SPDC). Three-level states are encoded in three modes of single photons where the modes are distinguished by a combination of their polarization and spatial degrees of freedom. We perform \robnew{bootstrap} tomography on a large set of preparations and a large set of measurements to obtain a GPT characterization of both. This bootstrap tomography approach differs from other experimental schemes that aim to characterize the geometry of three-level photonic systems, for example the one outlined in Ref. \cite{PhysRevLett.125.150401}, in that the correctness of quantum theory is not assumed in advance.
 
Our analysis uses a model selection technique in which data is collected for two independent runs of the experiment. The first data set is called the \textit{training set} and the second data set is called the \textit{test set}. For each of a set of candidate dimensions, we find the GPT characterization that best fits the training set. We then score these characterizations based on how well they predict the test set. This model selection technique was previously applied to the analysis of a quantum experiment in Daley {\em et al.}~\cite{daley2021experimentally}, where the goal was to adjudicate between different causal accounts of a Bell experiment. Here, it is applied to the problem of achieving \robnew{bootstrap} GPT tomography.

In the \robnew{original version of the bootstrap GPT tomography scheme, described in} Ref. \cite{Mazurek2017}, it is the Aka\"ike Information Criterion rather than a train-and-test technique that is used to adjudicate between models of different dimensions. Although the Aka\"ike Information Criterion has the advantage of yielding an estimate of the \robnew{probability that a given model in the slate of candidates minimizes the estimated information loss relative to the true model},
it does so only under strong assumptions about the error model underlying the experimental data, \robnew{and these assumptions do not, strictly speaking, hold in our experiment. Similar considerations hold for}
\edit{some of the proposed alternatives to the Aka\"ike Information Criterion (see, e.g.,~\cite{Schwarz,Hwang}).
An additional drawback of model selection techniques \robnew{based on the Aka\"ike Information Criterion and other criteria like it}
is that they are typically only reliable in situations where the number of data points is large compared to the number of parameters. \robnew{As this is not the situation in our experiment, we could not be confident of the verdicts delivered by such criteria.
The train-and-test technique, although more heuristic than those based on information criteria, can be applied
regardless of the error model and the relative number of data points and parameters.}
It is for these reasons that we have opted to use it here.
} 

\robnew{We now summarize the results obtained from our experimental data analysis and what can be inferred regarding possible deviations from quantum theory.  (For a more in-depth discussion of what one can infer about possible discrepancies in the dimension and shapes of the GPT state and effect spaces from experiments such as the one described here, we refer the reader to the introduction of Ref.~\cite{Mazurek2017}.)}

 There are two ways in which the true GPT describing the three-level system may differ from quantum-mechanical predictions. First, the experimental data might imply a deviation in the {\em dimension} of the GPT vector space (the number of parameters needed to specify a state or effect in the theory) relative to quantum theory. Second, the experimental data might imply no dimensional deviation, but a deviation in the {\em shape} of the state and effect spaces relative to quantum theory. 
     
 Because we are not assuming the correctness of quantum theory, the set of preparations and measurements that we implement might fail to be tomographically complete for the three-level system of interest. Furthermore, the additional preparations and measurements required to achieve tomographic completeness might only be implementable in a distinct experimental set-up from the one we consider, in particular, one involving exotic physics.  Because of this, a dimensional discrepancy between the true GPT and quantum theory might exist but be missed by our experiment.  
\rob{Nonetheless,}  if a dimensional discrepancy exists, it is 
 possible that it could be detected in a nonexotic experimental set-up (such as ours) because evidence for exotic physics can arise in conventional setups when these are probed at the precision frontier. In this sense, our experiment provides an opportunity for uncovering such a discrepancy. 
 
 \robnew{Applying our train-and-test model selection technique to the experimental data, we determine that the most likely dimension of the GPT vector space governing our three-level system is nine, in agreement with the dimension predicted by qutrit quantum theory.} 
 The fact that we find no evidence in favour of the hypothesis of a dimensional discrepancy implies that one of the following possibilities must hold:
 (i) it does not exist; (ii) it exists, but can only be detected in an experimental set-up involving exotic physics; (iii) it exists \rob{and can} be detected in the sort of experiment we have implemented \rob{but only} at higher precision than we achieved here.
 
Under the assumption that there is no dimensional discrepancy, we deduce \rob{from our experimental data} the geometry of the \rob{spaces of experimentally realized states and effects}, and from these we deduce the geometry of the \rob{spaces of logically consistent states and effects}. Together, these yield inner and outer bounding polytopes for the state and effect spaces of the true GPT governing the three-level system. \robnew{The bounds we obtain do not exclude the state and effect spaces of quantum theory and therefore our results are  consistent with the hypothesis that quantum theory is the true GPT underlying our experiment. Furthermore, using these bounding polytopes on the state and effect spaces, we are able to derive a quantitative bound on the extent to which the true GPT might violate the no-restriction hypothesis~\cite{Chiribella2010}. }
\robnew{ In Ref.~\cite{Mazurek2017}, this was done by estimating 
the ratio of the volumes of the inner and outer bounding polytopes for the state space.  In the higher-dimensional context considered here, however, computing these volumes is infeasible and so we instead estimate the ratio of the distance from the origin to the inner and outer bounding polytopes for the state space, averaged over a large number of directions.  This linear dimension ratio was found to be $0.80 \pm 0.04$.}



\robnew{In summary, our analysis is consistent with quantum theory being correct. If the true GPT {\em does} involve a dimensional deviation from quantum theory, 
then detecting it will require either achieving higher precision in a conventional set-up of the type we implemented, or an experimental set-up involving exotic physics.  Similarly, if the true GPT involves no dimensional deviation, but a deviation in the shapes of the state and effect spaces, then any future experiment establishing this will find such deviations to be limited in scope by the quantitative extent established here.  Our experiment and others of its kind thereby provide an empirical constraint---a kind of stress test---that can be applied to any novel theory that is proposed as a competitor to quantum theory. 
}


\section{Basics of the GPT Framework}
A GPT describes an experiment operationally by predicting the probability distribution over outcomes for every measurement given every preparation. To characterize the set of effects appearing in a GPT, it is sufficient to consider binary-outcome measurements, since an effect that appears in an $N$-outcome measurement also appears in the binary-outcome measurement that coarse-grains all the outcomes that are not associated to that effect. Furthermore, for any binary-outcome measurement, it is sufficient to specify the probability of only one of its outcomes, say the 0 outcome, since by normalization the probability of the other is then fixed. Thus, for each of a set of $m$ preparations $\{P_i\}_{i=1}^{m}$ and a set of $n$ binary-outcome measurements $\{M_j\}_{j=1}^{n}$, the GPT need only specify the probability of obtaining the 0 outcome, denoted $p(0|P_i,M_j)$. The full set of probabilities obtained from each of the preparation-measurement pairs in an experiment can be organized into an $m \times n$ matrix $D$, where each row represents a preparation, and 
each column represents a measurement:
\begin{equation*}
D = 
\begin{pmatrix}
p(0|P_1,M_1) & p(0|P_1,M_2) & \dots & p(0|P_1,M_n) \\
p(0|P_2,M_1) & p(0|P_2,M_2) & \dots & p(0|P_2,M_n) \\
\vdots & \vdots & \ddots & \vdots \\
p(0|P_m,M_1) & p(0|P_m,M_2) & \dots & p(0|P_m,M_n) \\
\end{pmatrix}.
\end{equation*}

The matrix $D$ encapsulates all of the GPT's predictions about the preparations and measurements in the experiment. If $D$ is rank $k$, then one can perform the decomposition $D = SE$, where $S$ is an $m \times k$ matrix and $E$ is a $k \times n$ matrix. The rows of $S$ are called \textit{GPT state vectors} and are denoted $\bm{s}_i, \ i \in \{1,2,\dots, m\}$. The columns of $E$ are called \textit{GPT effect vectors} and are denoted $\bm{e}_j,\ j \in \{1,2,\dots, n\}$. Therefore, each element of $D$ can be written as 
\begin{equation}
\label{prob}
    D_{i,j} = p(0|P_i,M_j) = \bm{s}_i \cdot \bm{e}_j.
\end{equation}

A GPT is fully specified by the set of all physically realizable state vectors, termed the \textit{GPT state space} and denoted $\bm{\mathcal{S}}$, as well as the set of all physically realizable effect vectors, termed the \textit{GPT effect space} and denoted $\bm{\mathcal{E}}$. The decomposition of $D$ is not unique. Specifically, for any invertible $k\times k$ matrix $\Lambda$, one can write $D = SE = S\Lambda \Lambda^{-1}E$. For each choice of $\Lambda$, the \rob{rows of the} matrix $S\Lambda$ consists of a set of valid GPT state vectors and the \rob{columns of the} matrix $\Lambda^{-1}E$ consists of a set of valid GPT effect vectors. For any two such decompositions, therefore, the GPT state and effect spaces defined by one are linearly related to those defined by the other.  There is consequently a linear freedom in the GPT state and effect spaces and \robn{stipulating a particular choice of linear transformation is therefore just a convention.}

GPTs are typically formulated such that any convex combination of preparation (measurement) procedures is also a valid preparation (measurement) procedure. This is justified by the requirement of including classical probability theory as a subtheory of the GPT, which allows for arbitrary mixing and post-processing. Consequently, the GPT state and effect spaces are convex. In addition, it is assumed that for any valid GPT effect vector $\bm{e}_j$ in the set $\bm{\mathcal{E}}$, the vector $\bar{\bm{e}}_j = \bm{u}-\bm{e}_j$ is also a valid effect vector, where  the vector $\bm{u}$ is called the unit effect and assigns probability 1 to all states.  We adopt the convention that $\bm{u} = (1,0,\dots,0)^T$, which implies that the first component of every normalized state must be 1, or equivalently, that the first column of the matrix $S$ contains only 1's.
We also adopt the convention that the effect with index $j=1$ is the unit effect, so that the first column of the matrix $D$ also contains only 1's. Thus, each $k$-dimensional state vector is fully described by a set of $k-1$ parameters.

We call a vector $\bm{s}$ a \textit{logically possible state} if it assigns a valid probability to every measurement effect allowed by the GPT. More formally, we define the space of logically possible states, denoted $\bm{\mathcal{S}}^{\text{logical}}$, as the set of states $\bm{s}$ such that $\forall \bm{e} \in \bm{\mathcal{E}}$ : $0 \leq \bm{s} \cdot \bm{e} \leq 1$ and $\bm{s}\cdot \bm{u} = 1$. From its definition, it can be shown that $\bm{\mathcal{S}}^{\text{logical}}$ is the intersection of the geometric dual of $\bm{\mathcal{E}}$ and the hyperplane $\bm{s}\cdot \bm{u} = 1$. For simplicity, we write $\bm{\mathcal{S}}^{\text{logical}} = \text{dual}(\bm{\mathcal{E}})$. In an analogous manner, we define the set of \textit{logically possible effects}, denoted $\bm{\mathcal{E}}^{\text{logical}}$, as the set of all vectors $\bm{e}$ such that $\forall \bm{s} \in \bm{\mathcal{S}}$ : $0 \leq \bm{s} \cdot \bm{e} \leq 1$. 
\rob{Again, for simplicity,} we write $\bm{\mathcal{E}}^{\text{logical}} = \text{dual}(\bm{\mathcal{S}})$.

GPTs in which $\bm{\mathcal{S}}^{\text{logical}} = \bm{\mathcal{S}}$ and $\bm{\mathcal{E}}^{\text{logical}} = \bm{\mathcal{E}}$ are said to satisfy the \textit{no-restriction hypothesis}~\cite{Chiribella2010}. For GPTs which satisfy the no-restriction hypothesis, all logically allowed GPT states are physically allowed and similarly for the effects. It is not possible for an experiment to prove that a particular GPT satisfies the no-restriction hypothesis exactly. This is because the number of states and effects characterized in an experiment will always be finite, and noise in their implementation will bound the realized GPT spaces away from the theoretical ideal. Therefore, the best one can do is constrain the extent to which any GPT that is logically consistent with the experimental data can violate the no-restriction hypothesis.

\section{The GPT State and Effect Spaces Describing a Qutrit}

One of the most familiar examples of a GPT is quantum mechanics itself. Here we will consider qutrit quantum mechanics.  In this case, the GPT state space $\bm{\mathcal{S}}^{\rm qutrit}$ is isomorphic to the set of all 
positive semi-definite trace-one linear operators acting on a 3-dimensional complex vector space,
i.e., the set ${\rm QStates}(\mathbb{C}^3) = \{\rho \in \mathcal{L}(\mathbb{C}^3) :  \rho\ge 0, {\rm Tr}(\rho)=1 
\}$. The GPT effect space 
$\bm{\mathcal{E}}^{\rm qutrit}$ is isomorphic to the set of all positive semi-definite linear operators with eigenvalues less than or equal to one, i.e., the set ${\rm QEffs}(\mathbb{C}^3) = \{ Q \in \mathcal{L}(\mathbb{C}^3): Q \ge 0,  Q\le I \}$.

A particularly useful convention to adopt for the GPT representation of the spaces of qutrit quantum states and quantum effects is to use the generalized Bloch representation. This representation is constructed from the identity operator together with a particular basis of \rob{the 8-dimensional vector space of traceless Hermitian operators}
 (which is orthonormal relative to the Hilbert-Schmidt inner product), namely, the basis consisting of the eight Gell-Mann matrices~\cite{Goyal_2016}.  These matrices generate the Lie algebra of the group SU(3), in analogy with how the Pauli matrices generate the Lie algebra of the group SU(2).  Denoting the Gell-Mann matrices by $\{\lambda_\alpha \}_{\alpha=1}^{8}$, the nine operators used in the Bloch representation are: 
\begin{equation*}
\lambda_0 = \begin{pmatrix}
1 & 0 & 0\\
0 & 1 & 0 \\
0 & 0 & 1
\end{pmatrix}, \,
\lambda_1 = \begin{pmatrix}
0 & 1 & 0\\
1 & 0 & 0 \\
0 & 0 & 0
\end{pmatrix}, \, \lambda_2 = \begin{pmatrix}
0 & -i & 0 \\
i & 0 & 0 \\
0 & 0 & 0
\end{pmatrix},
\end{equation*}
\begin{equation}
\label{gellmann}
\lambda_3 = \begin{pmatrix}
1 & 0 & 0\\
0 & -1 & 0 \\
0 & 0 & 0
\end{pmatrix}, \,
\lambda_4 = \begin{pmatrix}
0 & 0 & 1\\
0 & 0 & 0 \\
1 & 0 & 0
\end{pmatrix}, \, \lambda_5 = \begin{pmatrix}
0 & 0 & -i \\
0 & 0 & 0 \\
i & 0 & 0
\end{pmatrix},
\end{equation}
\begin{equation*}
\lambda_6 = \begin{pmatrix}
0 & 0 & 0\\
0 & 0 & 1 \\
0 & 1 & 0
\end{pmatrix}, \,
\lambda_7 = \begin{pmatrix}
0 & 0 & 0\\
0 & 0 & -i \\
0 & i & 0
\end{pmatrix}, \, \lambda_8 = \tfrac{1}{\sqrt{3}}\begin{pmatrix}
1 & 0 & 0\\
0 & 1 & 0 \\
0 & 0 & -2
\end{pmatrix}.
\end{equation*}
One easily verifies that the Gell-Mann matrices are Hermitian, traceless ($\text{Tr}[\lambda_\alpha] = 0  \ \forall \alpha \neq 0$), and orthonormal relative to the Hilbert-Schmidt inner product on the operator space ($\text{Tr}[\lambda_\alpha \lambda_\beta] = 2 \delta_{\alpha \beta} \ \forall \alpha,\beta \neq 0$, where the normalization is taken to be 2). With this notation, the Bloch representation of any \rob{state} $\rho\in {\rm QStates(\mathbb{C}^3)}$ and any \rob{effect} $Q \in {\rm QEffs(\mathbb{C}^3)}$ is given by $${\rho} = \frac{1}{3}{I}+\frac{1}{2}\sum_{\alpha=1}^8 s_\alpha {\lambda}_\alpha $$ and $${Q} = e_0{I} + \sum_{\alpha=1}^{8} e_\alpha {\lambda}_\alpha,$$ respectively. The GPT state vectors are given by the Bloch coefficients of $\rho$: $\bm{s} = (1, s_1, \dots, s_8)$, where $s_\alpha =\text{Tr}[\rho \lambda_\alpha]\; \rob{\forall \alpha \in \{0,\dots,8\}}$. Furthermore, the GPT effect vectors are given by the Bloch coefficients of $Q$: $\bm{e} = (e_0,e_1,\dots, e_8)$ where \rob{$e_\alpha = \frac{1}{2}\text{Tr}[Q \lambda_\alpha]\; \ \forall \alpha \in \{0,\dots,8\}$.}
 It follows that $p(0|P_i,M_j) = \text{Tr}[\rho_i Q_j] =  \bm{s}_i \cdot \bm{e}_j$, thereby matching Eq.~\eqref{prob}. Note that in this generalized Bloch representation, the quantum unit effect, $I$, is represented by the GPT effect vector having $e_0=1$ and $e_{\alpha}=0, \forall \alpha\ne 0$, so it abides by the convention that the unit effect vector is taken to be ${\bf u}=(1,0,\ldots,0)$.

Let $\tilde{\bm{s}} = (s_1, \dots, s_8)$ be the 8-dimensional vector obtained from $\bm{s}$ by eliminating its first component, and let $||\tilde{\bm{s}}|| = \sqrt{s_1^2 + s_2^2 + \dots + s_8^2}$. It has been shown that $\bm{s}$ is a GPT state vector for a valid qutrit state in the generalized Bloch representation \rob{($\bm{s}\in \bm{\mathcal{S}}^{\rm qutrit}$)} if and only if it satisfies the following constraints~\cite{KIMURA2003339}:
\begin{equation}
\label{constraints}
\begin{array}{ll}
           ||\tilde{\bm{s}}|| \leq \frac{2}{\sqrt{3}} \\
            \frac{2}{9} - \frac{1}{2}||\tilde{\bm{s}}||^2 + \frac{1}{2}\sum_{\alpha,\beta,\gamma} g_{\alpha \beta \gamma} s_\alpha s_\beta s_\gamma \geq 0,
        \end{array}
    \end{equation}
where  $g_{\alpha \beta \gamma}$ are the elements of the completely symmetric structure tensor of $SU(3)$. The elements of $g$ are defined implicitly
via the anti-commutation relations of the Gell-Mann matrices, \cite{KIMURA2003339, Bruning2011}
\begin{equation}
\{\lambda_\alpha, \lambda_\beta \} = \frac{4}{3}\delta_{\alpha \beta}I+2g_{\alpha \beta \gamma}\lambda_\gamma, \quad  \forall \alpha, \beta, \gamma \neq 0.
\end{equation}
The first constraint in Eq.~\eqref{constraints} corresponds to inclusion in an 8-dimensional hypersphere, which is the qutrit analog of the well-known Bloch sphere representation of qubits. The second constraint in Eq.~\eqref{constraints} has no qubit analog and makes the qutrit state space geometry much more complex.

Defining $\tilde{\bm{e}} = (e_1,\dots, e_8)$ and $||\tilde{\bm{e}}|| = \sqrt{e_1^2 + \cdots + e_8^2}$, we have that $\bm{e}$ is the GPT effect vector for a valid qutrit effect \rob{in the generalized Bloch representation} ($\bm{e} \in \bm{\mathcal{E}}^{\text{qutrit}}$) if and only if it satisfies the following constraints:
\begin{equation}
\label{constraintseffect}
\begin{array}{ll}
           0 \leq e_0 \leq 1 \\
           ||\tilde{\bm{e}}|| \leq \sqrt{3} \text{Min}\left(e_0, \ 1-e_0\right) \\
            e_0^{3} - e_0||\tilde{\bm{e}}||^2 + \frac{2}{3}\sum_{\alpha, \beta, \gamma} g_{\alpha \beta \gamma} e_\alpha e_\beta e_\gamma \geq 0 \\
            \begin{aligned}
            (1-e_0)^3 - (1 - e_0&)||\tilde{\bm{e}}||^2 - \\ & \frac{2}{3}\sum_{\alpha, \beta, \gamma} g_{\alpha \beta \gamma} e_\alpha e_\beta e_\gamma \geq 0.
            \end{aligned}
        \end{array}
    \end{equation}
The constraints in Eq. (\ref{constraintseffect}) are derived following the same steps as are followed to derive Eq.~\eqref{constraints}~\cite{KIMURA2003339}, with the added condition that $Q \leq I$. Note that the third and fourth conditions follow from $Q\ge0$ and $I -Q\geq 0$ respectively.

 To visualize the geometry of the \rob{9-dimensional} qutrit state \rob{and effect spaces},
   it is useful to consider the geometry of various 3-dimensional projections of these.
We follow the approach taken in Ref.~\cite{karol} to the study of the geometry of the qutrit state space
 (see also Refs.~\cite{Bengtsson2006,SZYMANSKI2018148,pirsa_16120010}), and adapt this approach to the case of the effect space.
  
 The \textit{joint numerical range} of a set of $k$ operators $\{A_\alpha \}_{\alpha=0}^{k-1}$ is defined as \cite{karol}
 \begin{equation}\label{JNRdefn}
 \begin{split}
W(A_0,\ldots, A_{k-1}) = \{\bm{w} \in \mathbb{R}^k : w_\alpha  = \langle & \psi| A_\alpha |\psi\rangle ,
\\
& \forall |\psi\rangle \in 
\mathbb{C}^{d} 
     \},
\end{split}
 \end{equation} 
where $\mathbb{C}^{d}$ is the complex vector space of dimension $d$.
Consider the nine operators defined by 
$$A_0=I,\;\; A_{\alpha} =\lambda_{\alpha},\; \forall \alpha \in \{1,\dots, 8\},$$
i.e., the identity matrix and the eight Gell-Mann matrices, defined in Eq.~\eqref{gellmann}.  The joint numerical range $W(A_0,\ldots, A_8)$  is equal to the set of 9-dimensional generalized Bloch vectors associated to pure qutrit states (i.e., the Bloch vectors including the $s_0$ component),
 \begin{align}
\bm{\mathcal{S}}_{\rm pure}^{\rm qutrit}&= W(A_0,\ldots, A_8).
 \end{align} 
It follows that the {\em convex hull} of $W(A_0,\ldots, A_8)$  is equal to $\bm{\mathcal{S}}^{\rm qutrit}$, which corresponds to the set of 9-dimensional generalized Bloch vectors associated to {\em all} qutrit states, characterized in Eq.~\eqref{constraints},
\begin{align}
 \begin{split}
\bm{\mathcal{S}}^{\rm qutrit}&=  \{\bm{s} \in \mathbb{R}^9 : s_\alpha = \text{Tr}[\rho A_\alpha]\ \rob{\forall \alpha \in \{0,\dots,8\}} , \  \\ & \qquad \qquad \qquad \qquad \qquad \forall \rho \in {\rm QStates}(\mathbb{C}_3)  \}, 
\end{split} \nonumber\\
&= {\rm cvxhull} \left[W(A_0,\ldots, A_8)\right].
 \end{align} 
For some arbitrary triple of these nine operators, $\{ A_{\mu}, A_{\nu}, A_{\omega} \}$, the convex hull of their joint numerical range describes the {\em projection} of $\bm{\mathcal{S}}^{\rm qutrit}$ into the 3-dimensional operator subspace spanned by $\{ A_{\mu}, A_{\nu}, A_{\omega}\}$,
\begin{align}\label{StateProjectionByJNR}
 \begin{split}
\mathcal{P}_{\mu \nu \omega}(\bm{\mathcal{S}}^{\rm qutrit})&= \{\bm{s} \in \mathbb{R}^3 : s_\alpha = \text{Tr}[\rho A_\alpha] , \rob{\forall \alpha \in \{\mu,\nu,\omega\},} \\ & \qquad \qquad \qquad \qquad \qquad \forall \rho \in {\rm QStates}(\mathbb{C}^3)  \}\end{split}  \nonumber\\
& = {\rm cvxhull} \left[W(A_{\mu}, A_{\nu}, A_{\omega})\right].
\end{align}

Similar considerations hold for the effects.  Taking $$B_0=\frac{1}{3}I , \;\; B_{\alpha} =\frac{1}{2}\lambda_{\alpha},\; \forall \alpha \in \{1,\dots, 8\},$$ the joint numerical range $W(B_0, \ldots, B_8)$  is equal to the set of generalized Bloch vectors associated to {\em atomic} effects on qutrits. An atomic quantum effect is one described by a rank-1 projector, so that the set of atomic quantum effects for a qutrit is simply ${\rm AtmQEffs} (\mathbb{C}^3) = \{ |\psi\rangle \langle \psi| : |\psi\rangle \in \mathbb{C}^3\}$. Denoting the set of GPT effect vectors that are associated to atomic qutrit effects by $  \bm{\mathcal{E}}_{\rm atomic}^{\rm qutrit}$, we have
   
   \begin{align}
  \bm{\mathcal{E}}_{\rm atomic}^{\rm qutrit}&= W(B_0, \ldots, B_8).
 \end{align} 
  For all such GPT effect vectors, $e_0 = {\rm Tr}(\tfrac{1}{3} |\psi\rangle\langle\psi|) =\tfrac{1}{3}$.

To transition from the set of atomic quantum effects to the set of {\em all} quantum effects, the following procedure suffices.  For each atomic quantum effect $Q \in {\rm AtmQEffs}(\mathbb{C}^3)$, add to the set its complement, $I-Q$. In addition, add to the set the unit effect, $I$, and the zero effect, $0$, which assigns probability 0 when paired with all states. Finally, close under convex combinations.  In summary, if one denotes the set of complements of atomic quantum effects by $\overline{{\rm AtmQEff}}(\mathbb{C}^3)$, then the full set of quantum effects satisfies
   \begin{align}
   \begin{split}
 {\rm QEff}(\mathbb{C}^3) = {\rm cvxhull}\big[  \{ 0\} \cup \{I\} \cup & {\rm AtmQEff}(\mathbb{C}^3) \\
  & \cup \overline{{\rm AtmQEff}}(\mathbb{C}^3)\big]
 \end{split}
   \end{align}
   
The same procedure allows one to go from the set of GPT effect vectors describing atomic qutrit effects in the generalized Bloch representation to the set of GPT effect vectors describing generic qutrit effects. Recall that ${\bf u}=(1,0,\ldots,0)$ denotes the GPT effect vector associated to the unit quantum effect. Letting  ${\bf 0}=(0,\ldots, 0)$ denote the GPT effect vector associated to the zero quantum effect and letting $\bar{\bm{\mathcal{E}}}_{\rm atomic}^{\rm qutrit}$ denote the complement of the set of atomic effects, $\bar{\bm{\mathcal{E}}}_{\rm atomic}^{\rm qutrit}  = \{ {\bf u}- {\bf e} : {\bf e} \in \bm{\mathcal{E}}_{\rm atomic}^{\rm qutrit}\}$, we have
\begin{align} \label{qutritspace}
\begin{split}
\bm{\mathcal{E}}^{\rm qutrit} 
&=  \{ \bm{e} \in \mathbb{R}^9 : e_\alpha = \text{Tr}[Q B_\alpha]\; \rob{\forall \alpha \in \{0,\dots,8\},} \\
& \qquad \qquad \qquad \qquad \qquad \qquad \forall Q \in {\rm QEff}(\mathbb{C}^3) \}
\end{split} \nonumber\\
&= {\rm cvxhull} \left[ \{{\bf 0 }\} \cup \{{\bf u}\} \cup \bm{\mathcal{E}}_{\rm atomic}^{\rm qutrit} \cup \bar{\bm{\mathcal{E}}}_{\rm atomic}^{\rm qutrit}\right].
\end{align} 

As with the states,  we wish to consider  the {\em projection} of $\bm{\mathcal{E}}^{\rm qutrit}$ into the 3-dimensional operator subspace spanned by $\{ B_{\mu}, B_{\nu}, B_{\omega}\}$ for various triples of operators.  Some preliminary definitions are useful for describing this projection.  Let ${\bf 0}^{(3)} \equiv (0,0,0)$; this is the 3-dimensional projection of ${\bf 0}$ for any choice of $\mu$, $\nu$, $\omega$. Let ${\bf u}^{(\mu \nu\omega)}$ denote the projection of ${\bf u}$  into the appropriate 3-dimensional subspace. Note that if $0\in \{\mu,\nu,\omega\}$, then the corresponding component of ${\bf u}^{(\mu \nu\omega)}$ is 1, with the other components being 0, while if $0\not\in \{\mu,\nu,\omega\}$, then ${\bf u}^{(\mu \nu\omega)}=\rob{{\bf 0}^{(3)}}$.
Finally, let $\overline{W}(B_{\mu}, B_{\nu}, B_{\omega})$ denote the complement of $W(B_{\mu}, B_{\nu}, B_{\omega})$ relative to ${\bf u}_{\mu \nu\omega}$, i.e., $\overline{W}(B_{\mu}, B_{\nu}, B_{\omega}) =  \{ {\bf u}^{(\mu \nu\omega)}- {\bf e} : {\bf e} \in W(B_{\mu}, B_{\nu}, B_{\omega})\}$. For those projections where  ${\bf u}^{(\mu \nu\omega)}=\rob{{\bf 0}^{(3)}}$, $\overline{W}(B_{\mu}, B_{\nu}, B_{\omega})$ is simply the inversion about the origin of $W(B_{\mu}, B_{\nu}, B_{\omega})$.

Relative to these definitions, the 3-dimensional projections of $\bm{\mathcal{E}}^{\rm qutrit}$ are given by:
\begin{align}\label{EffProjectionByJNR}
\mathcal{P}_{\mu \nu \omega}(\bm{\mathcal{E}}^{\rm qutrit})\nonumber &=  \{\bm{e} \in \mathbb{R}^3 : e_\alpha = \text{Tr}[Q B_\alpha] , \forall Q \in {\rm QEff}(\mathbb{C}^3)  \} \nonumber \\
\begin{split}
& = {\rm cvxhull} \big[\{{\bf 0}^{(3)}\} \cup \{{\bf u}^{(\mu \nu\omega)}\} \\ & \qquad  \cup  W(B_{\mu}, B_{\nu}, B_{\omega}) \cup \overline{W}(B_{\mu}, B_{\nu}, B_{\omega}) \big].
 \end{split}
 \end{align} 


\begin{figure*}[hbt!]
     \centering
     \includegraphics[width=\textwidth,keepaspectratio]{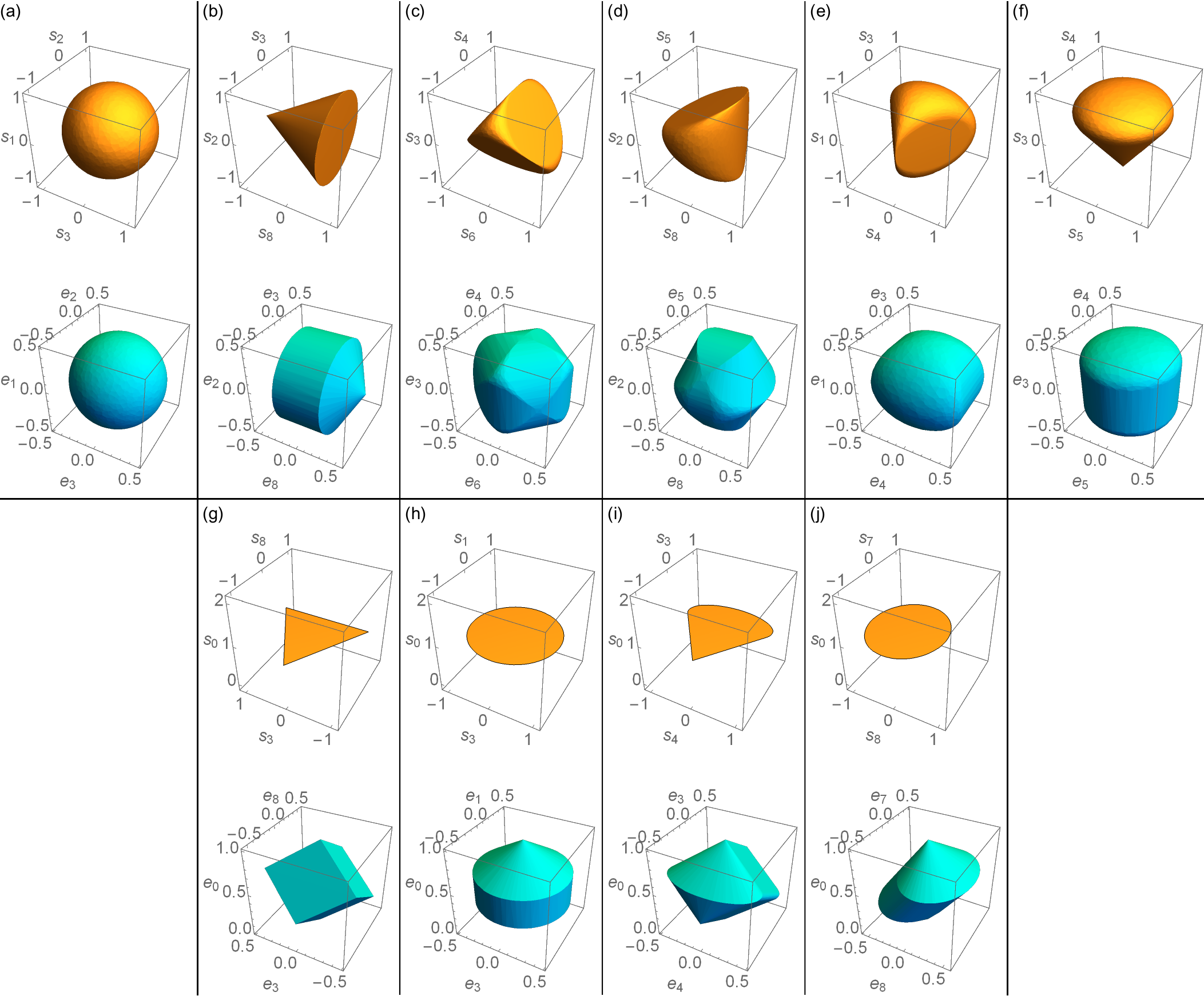}
     \caption{\edit{A sampling of 3-dimensional projections of the 9-dimensional qutrit state space (top shape of each subfigure, displayed in yellow) and of the 9-dimensional qutrit effect space (bottom shape of each subfigure, displayed in cyan)}. 
 For each subfigure (a)-(j), we specify the triple of indices $(\mu, \nu, \omega)$ describing which 3-dimensional projection is plotted, as well as the type of the state space projection.  In the cases (a)-(f), the type is a pair of numbers $(l,f)$, counting the flat segments and flat faces in the 2-dimensional boundary, while in the cases (g)-(f), it is a single number $l$ counting the flat segments in the 1-dimensional boundary.   (a) $(\mu,\nu,\omega)=(1,2,3)$,  $(l,f)=(0,0)$; (b) $(\mu,\nu,\omega)=(2,3,8)$, $(l,f)=(\infty,1)$; (c) $(\mu,\nu,\omega)=(3,4,6)$, $(l,f)=(0,4)$; (d) $(\mu,\nu,\omega)=(1,3,4)$, $(l,f)=(0,2)$; (e) $(\mu,\nu,\omega)=(2,5,8)$, $(l,f)=(1,2)$; (f) $(\mu,\nu,\omega)=(3,4,5)$, $(l,f)=(\infty,0)$; (g) $(\mu,\nu,\omega)=(0,3,8)$, $l = 3$; (h) $(\mu,\nu,\omega)=(0,1,3)$, $l = 0$; (i) $(\mu,\nu,\omega)=(0,3,4)$, $l = 2$; (j) $(\mu,\nu,\omega)=(0,7,8)$, $l = 0$.
  }
     \label{fig:maxextentstate}
 \end{figure*}

Fig. \ref{fig:maxextentstate} shows plots of a number of 3-dimensional projections of the qutrit state space (yellow shapes) and the qutrit effect space (cyan shapes).  Before describing some of their features, we briefly explain how these plots are generated. 

To plot a given 3-dimensional projection of $\bm{\mathcal{S}}^{\rm qutrit}$, we begin by generating a large quantity of random GPT state vectors \robn{that are on the boundary of the set defined by} 
Eq.~(\ref{constraints})~\footnote{We draw the set of random qutrit quantum states and effects from the SU(3) invarient Haar distribution. We discuss this method in section \ref{setup}}. For every GPT state vector in our random set, we project it down into the 3-dimensional space of interest. We then plot the convex hull of these 3-dimensional vectors.  Similarly, to generate a plot of any given 3-dimensional projection of $\bm{\mathcal{E}}^{\rm qutrit}$, we generate a large quantity of random GPT effect vectors \robn{that are on the boundary of the set defined by}
Eq.~\eqref{constraintseffect} and with $e_0 = \tfrac{1}{3}$, which is to say, vectors associated to \mike{atomic} qutrit effects.  For every GPT effect vector obtained in this way, we compute its complement effect. We then project both the atomic effects and their complements down into the 3-dimensional space of interest. Finally, we plot the convex hull of the set of \rob{projected} GPT effect vectors, their \rob{projected} complements, and the projected zero and unit effect vectors. As the number of vectors defining these convex shapes is always finite, the shapes depicted in the plots are polytopes.  Indeed, a close examination of Fig. \ref{fig:maxextentstate} reveals that they are not smooth, in contrast to the true 3-dimensional projections.  Nonetheless, because the sampling is quite dense, the plots provide a good visual approximation to the true shapes.

Consider first the 3-dimensional projections of the 9-dimensional qutrit state space onto a triple of Gell-Mann matrices,
   that is, the projections
   where $0 \notin \{\mu,\nu,\omega\}$.  There are ${8}\choose{3}$ =$56$ of these. In Ref.~\cite{pirsa_16120010,SZYMANSKI2018148}, the connection to joint numerical ranges was leveraged to show that all 3-dimensional projections of the qutrit state space
    can be classified into different types 
    according to the number of {\em flat segments} in their \rob{2-dimensional} boundary and the number of {\em flat faces} in their  \rob{2-dimensional} boundary.
A given type of 3-dimensional projection, therefore, can be specified by a pair of natural numbers, denoted $(l,f)$, where $l$ is the number of flat segments and $f$ is the number of flat faces.  

Fig.~\ref{fig:maxextentstate}(a)-(f) depicts a sampling of 3-dimensional projections of the qutrit state space, corresponding to six distinct types.  The projection depicted in (a), for example, is of type $(l,f)=(0,0)$ because it is a ball and therefore has no flat segments or flat faces, while for (e), the type is $(l,f)=(1,2)$ because it is has two flat faces (one on the top and one on the bottom) and a single flat segment (connecting the two flat faces \rob{at the points where they are furthest apart}).

In Fig.~\ref{fig:maxextentstate}(g)-(j), we display a sampling of 3-dimensional projections of the qutrit state space where $0 \in \{ \mu,\nu,\omega \}$, 
so that the component $s_0$ is included in the projection.  There are ${8}\choose{2}$ =$28$ of these projections.  By convention, we plot the $s_0$ component along the vertical axis.   Because the states are normalized, their projections are confined to the $s_0=1$ plane and the shape is only nontrivial along the other two axes.  It follows that the $s_0=1$ section of the projection onto the 3-dimensional  subspace $(0\nu\omega)$ is equivalent to the  projection onto the 2-dimensional subspace $(\nu\omega)$.
All 2-dimensional projections of the qutrit state space can be classified into different types in a manner analogous to the 3-dimensional projections, namely, by the number of flat segments in their \rob{1-dimensional} boundary.  We denote this number by $l$.   
For example, the projection in Fig. \ref{fig:maxextentstate}(g) is of type $l = 3$ because it has three flat segments in its boundary.  Note that the projections depicted in (h) and (j) are of the same type, $l=0$, because they both have no flat segments in their boundary. 
 This pair of examples illustrates the fact that  two projections of the same type can still have distinct shapes. 

   Although we do not present an exhaustive set of projection types
     in Fig. \ref{fig:maxextentstate}, our sampling is sufficient to give a sense of the 
     geometric features that are characteristic of the qutrit state space.

We now turn to the projections of $\bm{\mathcal{E}}^{\rm qutrit}$. In Fig.~\ref{fig:maxextentstate}, we display projections of $\bm{\mathcal{E}}^{\rm qutrit}$ alongside the corresponding projection
 of $\bm{\mathcal{S}}^{\rm qutrit}$.
As before, therefore, projections where $0 \not\in \{\mu,\nu,\omega\}$ appear in Fig.~\ref{fig:maxextentstate}(a)-(f), while projections where $0 \in \{\mu,\nu,\omega\}$ (i.e., which include the $e_0$ component) appear in Fig.~\ref{fig:maxextentstate}(g)-(j).
Given that for effects, unlike states, the 
 component $e_0$ can be any value in the interval  $[0,1]$,
even the projections where $0 \in \{\mu,\nu,\omega\}$
are nontrivial along all three axes.

It is useful to take note of some general features of the shapes appearing in Fig.~\ref{fig:maxextentstate}. The set of trace-one quantum effects is always precisely the same as the set of normalized quantum states, ${\rm cvxhull} \left({\rm AtmQEffs}(\mathbb{C}^d)\right)= {\rm QStates}(\mathbb{C}^d)$. It follows that the $e_0=1/3$ section of the space of GPT effect vectors for a qutrit (which describes the set of trace-one quantum effects in the generalized Bloch representation we are considering) should be of precisely the same shape as the $s_0=1$ section of the space of GPT states for a qutrit.
 \rob{This also holds true} for every 3-dimensional projection \rob{where $0\in \{\mu,\nu,\omega \}$}:
  the $e_0=1/3$ section of the projected effect space should coincide with the $s_0=1$ section of the projected state space.
  This is precisely what is seen in Figs.~\ref{fig:maxextentstate}(g)-(j). Furthermore, in those 3-dimensional projections where \rob{$0\not\in \{\mu,\nu,\omega\}$},
   we expect the projection of the effect space to be symmetric under inversion about the origin, for the reasons discussed in the paragraph below Eq.~\eqref{qutritspace}.  This feature is also evident in Figs.~\ref{fig:maxextentstate}(a)-(f).

The 3-dimensional projection depicted in Fig.~\ref{fig:maxextentstate}(g) is particularly \mike{intuitive} to interpret.  It describes a classical three-level system, i.e., a classical trit.  This projection is associated to the triple of Gell-Mann matrices $\lambda_0$, $\lambda_3$, and $\lambda_8$, which form a commutative algebra.  It follows that the joint eigenstates of these three Gell-Mann matrices describe a triple of perfectly distinguishable states, and the associated projection of the qutrit state space is the convex hull of these, i.e., a triangle, which is the state space of a classical trit.  Meanwhile, the associated projection of the qutrit effect space is the corresponding three-dimensional hypercube of classical effects~\footnote{This is the qutrit analogue of how the projection of the space of {\em qubit} states into any 2-dimensional subspace which includes the identity yields a line segment state space, which is the state space of a classical bit, while the corresponding projection of the space of qubit effects yields a diamond-shaped effect space, which is the effect space of a classical bit.}.

 Regarding the precise shapes of the other projections depicted in Fig.~\ref{fig:maxextentstate}, we restrict ourselves to a few observations.  
 
 The fact that Fig.~\ref{fig:maxextentstate}(a) depicts shapes that are isomorphic to a Bloch ball of states and a Bloch ball of trace-one effects follows from the fact that the algebra defined by the four operators $\{ \tfrac{2}{3}\lambda_0 +\tfrac{1}{\sqrt{3}}\lambda_8, \lambda_1, \lambda_2, \lambda_3\}$ is\mike{, in fact,} the Pauli algebra, and so describes a qubit embedded in the qutrit.   
 
 Meanwhile, the three operators $\{ \tfrac{2}{3}\lambda_0 +\tfrac{1}{\sqrt{3}}\lambda_8, \lambda_1, \lambda_3\}$ describe the Pauli algebra without the Pauli-$Y$, and so are associated to a real-amplitude qubit (or `rebit'). The state space projection depicted in Fig.~\ref{fig:maxextentstate}(h), therefore, is the Bloch disc associated to a rebit. The effect space projection depicted in Fig.~\ref{fig:maxextentstate}(h), meanwhile, is a bit more complicated.  It is the convex hull of the trace-one rebit effects (the Bloch disc at $e_0=1/3$), together with their complements relative to the qutrit unit effect (the trace-two rebit effects, depicted as a disc at $e_0=2/3$), as well as the zero effect and the qutrit unit effect.  
 
 More generally, some understanding of the shapes of the $(\mu,\nu,\omega)$ projections (of both the state and effect spaces) can typically be gleaned by considering the algebra defined by the operators $\lambda_{\mu}, \lambda_{\nu}$, and $ \lambda_{\omega}$ and how this algebra embeds in the full qutrit algebra.  

Quantum theory satisfies the no-restriction hypothesis, so the set of logically possible qutrit states is precisely equal to the set of physically realizable qutrit states and the set of logically possible qutrit effects is precisely equal to the set of realizable qutrit effects. If the GPT describing the three-level system of our experiment were to differ from quantum theory such that it {\em failed} to satisfy the no-restriction hypothesis, then this would manifest as a gap between the physically realizable and logically possible state spaces (equivalently, a gap between the physically realizable and logically possible effect spaces). Any such gap would also be manifest in some of the 3-dimensional projections of these.

\section{Description of the Experiment}
\label{setup}

 \begin{figure*}[hbt!]
     \centering
     \includegraphics[width=\textwidth,keepaspectratio]{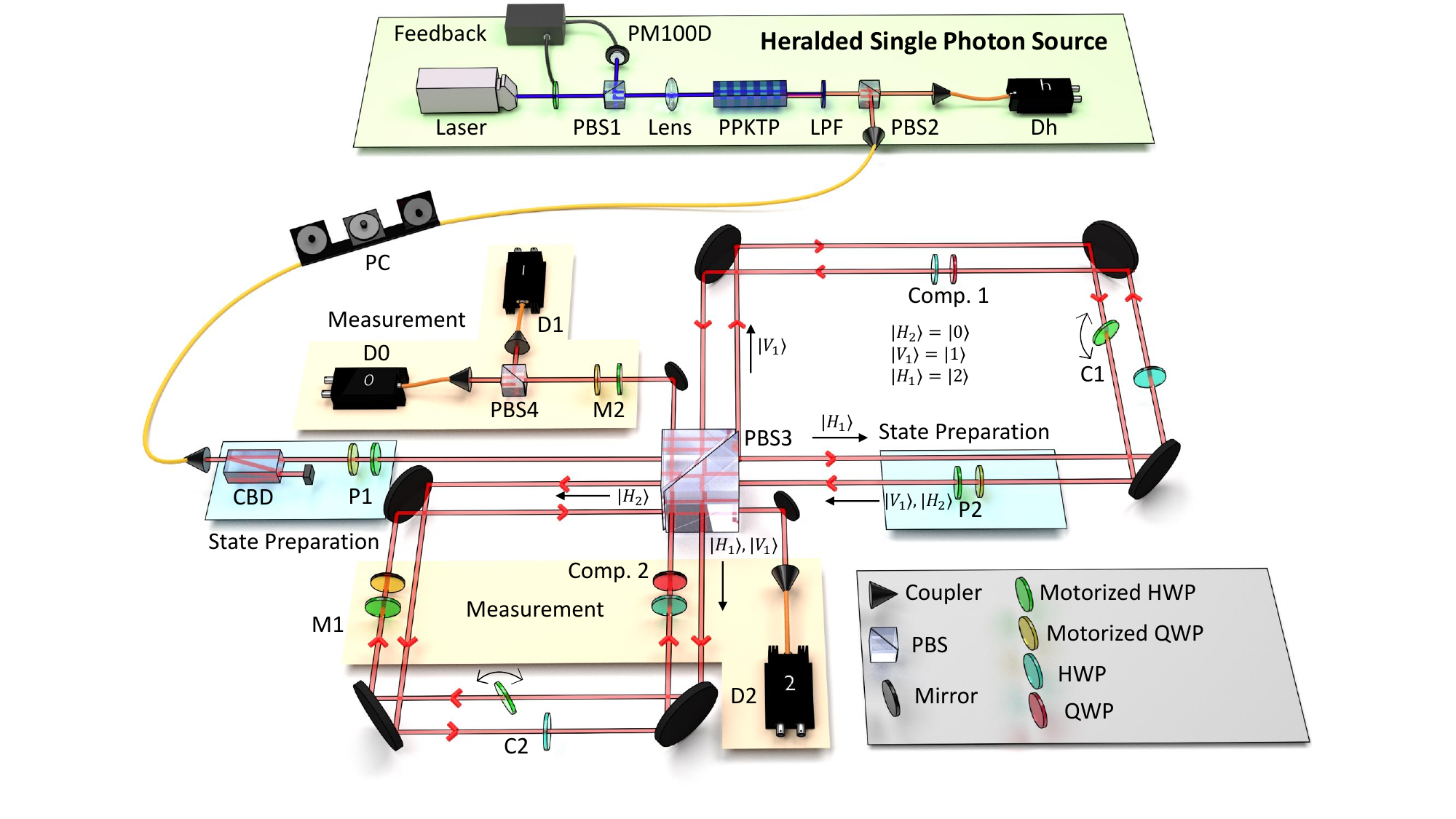}
     \caption{Single photons at $808 \, \mathrm{nm}$ are generated via type-II SPDC using a continuous wave laser diode at $404\, \mathrm{nm}$ focused onto a PPKTP crystal with a lens. Polarization drifting effects of the source are actively compensated using a power meter (PM100D) connected via a feedback loop to a motorized HWP. Excess 404nm light is filtered out with a long pass filter (LPF). Photons detected at APD $Dh$ are used to herald the arrival of photons arriving at APDs $D0$ (outcome 0), $D1$ (outcome 1), and $D2$ (outcome 2) through the displaced sagnac interferometer with a coincidence window of $3 \, \mathrm{ns}$. Three-level states are prepared in the modes $|H_1\rangle$, $|V_1\rangle$, and $|H_2\rangle$ using QWP-HWP pairs P1, P2 (highlighted in blue) and are measured using HWP-QWP pairs M1, M2 (highlighted in orange). \edit{The interference visibility of each interferometer is actively compensated using two HWP pairs $C1$ and $C2$. One HWP of each pair is placed on a pivoting motorized mount, allowing for fine control of the relative optical path length. The non-motorized wave plates, labeled as Comp. 1 and Comp. 2 and shown in blue and red, are used to balance the optical path lengths of the interferometer arms and always have optical axis oriented along the horizontal.} The arrows shown within the interferometer arms are displayed for clarity of photon direction.}
     \label{fig:setup}
 \end{figure*}

Our experimental setup is shown in Fig. \ref{fig:setup}. A heralded single photon source uses a continuous-wave diode laser with output at $404 \, \mathrm{nm}$ that is focused with a lens onto a $15 \, \mathrm{mm}$ PPKTP crystal to perform Type-II SPDC. Excess $404 \, \mathrm{nm}$ light is filtered out with a long pass filter (LPF), and resulting $808 \, \mathrm{nm}$ photons are split using a polarizing beam splitter (PBS2). The transmitted photons are received by an avalanche photo diode (APD), denoted $Dh$, which heralds the arrival of the reflected photons within a coincidence window of $3 \, \mathrm{ns}$. To control polarization drifting effects characteristic to the laser, the reflected port of PBS1 is actively monitored with a power meter (PM100D) and is connected via a feedback loop to a motorized half wave plate (HWP) to ensure optimal crystal pump polarization at all times.

Photons from the source are coupled into a single-mode fiber with polarization control (PC) and are passed to an interferometer in a displaced Sagnac configuration. A calcite beam displacer (CBD) is used with the fiber PC to ensure vertically polarized photons are incident to the interferometer. Different states of the three-level system are prepared using motorized quarter wave plate (QWP) and HWP pairs P1 and P2 (highlighted in blue in Fig. \ref{fig:setup}). Preparation stage P1 is used to prepare a two-level state with modes $|H_1\rangle$ and $|V_1 \rangle$. These modes are split on PBS3. Mode $|H_1\rangle$ travels counterclockwise around the first interferometer loop and returns to PBS3 unchanged. Mode $|V_1\rangle$ travels clockwise around the first interferometer loop and is addressed by preparation stage P2 where the third mode of the three-level state, denoted $|H_2\rangle$,  is prepared. Each three-level state is fully constructed once all three modes recombine on PBS3. \rob{The measurement is varied}
 using motorized HWP-QWP pairs M1 and M2 (highlighted in orange in Fig. \ref{fig:setup}) and \rob{the photon is} detected at one of three APDs $D0$, $D1$, or $D2$, corresponding to a three-outcome measurement.  In our analysis, outcomes $1$ and $2$ are coarse-grained to give a binary-outcome measurement. \edit{The non-motorized QWP-HWP pairs in Fig. \ref{fig:setup} (shown in blue and red and labeled Comp. 1 and Comp. 2) are used to balance the optical path lengths of the interferometer arms and always have optical axis oriented along the horizontal}.

The beams in the interferometer arms are displaced to about $5 \, \mathrm{mm}$. All wave plates within the interferometer arms have $3.5 \, \mathrm{mm}$ holes allowing one beam to pass freely through. The interferometer is inside a box to reduce air currents and improve long term stability. \edit{Relative phases of the interferometer arms are controlled using tilted HWPs placed on pivoting motors (displayed skewed in wave plate pairs C1 and C2). The interference visibility of both interferometers is automatically checked and corrected every 20 minutes as it was noted that interferometer phase began to drift after approximately 30 minutes. The compensation scheme is as follows: For a specific experimental configuration in which maximum interference should occur on PBS3 and in which the count rate at detector D1 should be minimal, the motorized HWPs in either C1 or C2 (depending on the interferometer being compensated) is pivoted until a threshold count rate is achieved at detector D1, which corresponds to an optimal interference visibility. We obtain an optimal interference visibility for the first interferometer (top in Fig. \ref{fig:setup}) of $0.972$ and an optimal interference visibility for the second interferometer (bottom in Fig. \ref{fig:setup}) of $0.984$. The discrepancy between the two optimal visibility values is explained by the fact that the second interferometer has a shorter optical path length than that of the first, making it intrinsically more stable.}

To choose our experimental preparation settings, we first generate a number of random qutrit quantum states that are distributed according to the $SU(3)$ invariant Haar measure. This is accomplished by sampling a number of $3\times 3$ unitary matrices according to the Haar distribution, whose first columns are then taken to be a set of random pure quantum states \cite{Emerson2003, Mezzadri2006}. We then compute the HWP and QWP angles necessary to prepare each of these random states in our experimental setup and use these angles to set the QWP-HWP pairs P1 and P2 shown in Fig. \ref{fig:setup}. 

Every rank-1 projective effect $Q_j = |\psi_j\rangle \langle \psi_j|$ can be matched with the state to which it responds with unit probability, \rob{namely, $\rho_j = |\psi_j\rangle \langle \psi_j|$, so that}
 $\text{Tr}[\rho_j Q_j] =  \text{Tr}[|\psi_j\rangle \langle \psi_j |\psi_j\rangle \langle \psi_j|] = 1$. We choose our set of measurements such that $\{Q_j, \bar{Q}_j\} = \{|\psi_j\rangle \langle \psi_j|, I-|\psi_j\rangle \langle \psi_j|\}$, where each $|\psi_j\rangle \langle \psi_j|$ corresponds to one of the sampled preparation states. Here, $Q_j$ is the effect corresponding to outcome 0 and $\bar{Q}_j$ is the effect corresponding to the coarse-graining of outcomes 1 and 2. Therefore, the measurement settings are the wave plate angles \rob{that must be passed to the HWP-QWP pairs M1 and M2 shown in Fig. \ref{fig:setup} in order to}
   implement the \rob{$n$ projective measurements of the form $\{Q_j, \bar{Q}_j\}$}.
   We convert our three-outcome measurement into a binary-outcome measurement in post-processing. During the experiment however, we measure photon counts at all three detectors independently. 
\rob{There are a continuum of  experimental configurations such that coarse-graining outcomes 1 and 2 yields the binary-outcome measurement $\{Q_j, \bar{Q}_j\}$.  Every decomposition of the rank-2 projector $\bar{Q}_j$ into a pair of rank-1 projectors yields a distinct such configuration.}
However, ensuring that the effect $Q_j$ is implemented at detector D0 provides us with sufficient information to conduct our GPT analysis scheme. We therefore \rob{choose}
 {\em any} set of wave plate angles that \rob{associates the effect $Q_j$ with}
 detector D0, without imposing any constraints on the effects \rob{associated to}
  detectors D1 and D2. 
  
  Our choices of measurement settings produce many instances in which $p(0|P_i,M_j) = 1$ in theory. However, in practice, limitations of the experimental setup always bound the probabilities obtained  away from 1, so that the uncertainties on our counts never produce super-normalized outcomes. Note that although our choices of experimental preparations and measurements are informed by quantum theory, we do not assume its correctness in any of our analysis techniques.

For our first experiment, we choose $m = 100$ Haar-distributed preparation settings and $n = 100$ corresponding measurement settings for a total of $mn = 10^4$ experimental configurations. \edit{With a laser power of $0.5 \ mW$ incident to the PPKTP crystal, we report a second order correlation function value of $g^{(2)} = 0.0034$ and detect photon coincidences at a rate of approximately $2000 \, \mathrm{counts/s}$. We integrate each measurement configuration for $2 \, \mathrm{s}$.} Including the time it takes for the motorized wave plates to re-position between settings, and the time it takes to actively correct for interferometer phase drift, each run of the experiment takes approximately $16 \,\mathrm{h}$ to complete.

\section{Bootstrap Tomography in the GPT Framework}

\subsection{Inferring best-fit probabilities from experimental data}

We now discuss how to \rob{estimate}
 the probability matrix $D$ directly from experimental data. Heralded photon counts are detected at $D0$, $D1$, and $D2$. We assume Poissonian uncertainty on the counts obtained at each detector. One can obtain \rob{estimates of}
  the probabilities $p(0|P_i,M_j)$  by dividing the number of counts detected at $D0$ by the total number of counts obtained at all three detectors. We denote these relative frequencies by $f(0|P_i,M_j)$ and \rob{from them we} construct a frequency matrix $F$ that is an \rob{estimate of}
   the rank-$k$ probability matrix $D$.
     The experiment was conducted twice (back to back) using the same preparation and measurement settings. We call the first data set obtained the \textit{training set} and the second data set obtained the \textit{test set}. The training set, with frequency matrix $F^{\text{train}}$, is used to find the best-fit probability matrix $D$ that \rob{characterizes}
      the GPT underlying the experiment. Since the \rob{frequencies in the} matrix $F^{\text{train}}$ \rob{fluctuate away from the true probabilities}, 
$F^{\text{train}}$ will tend to be full rank, regardless of the rank of the underlying matrix $D$. Therefore, the problem at hand is to find, for each of a set of candidate ranks, the matrix $D$ of that rank that best fits the matrix of frequencies $F^{\text{train}}$. Specifically, the task can be formulated as the weighted low rank approximation problem
\begin{equation}
\label{opt}
\begin{aligned}
\text{minimize} \quad & \chi_k^{2 \ \text{train}} = \sum_{i=1}^{m} \sum_{j=1}^{n} \left(\frac{F_{ij}^{\text{train}}-D_{ij}}{\Delta F_{ij}^{\text{train}}}\right)^2, \\
\textrm{subject to} \quad & \text{rank}(D) = k\\
  & 0 \leq D_{ij} \leq 1    \\
\end{aligned}
\end{equation}
where $\Delta F_{ij}^{\text{train}}$ is the statistical uncertainty in $F_{ij}^{\text{train}}$.

Eq. (\ref{opt}) can be solved using an algorithm based on an alternating least squares approach \cite{Mazurek2017}. The column of ones corresponding to the unit effect that we include in $F$ are added in by hand and have uncertainty 0. Because $D$ is defined as the matrix that minimizes $\chi_k^{2 \ \text{train}}$, the entries in the same column of $D$ must also be exactly 1. (If this was not the case, then $\chi_k^{2 \ \text{train}}$ would be undefined for those entries.) For each of the candidate ranks $k$, we compute the best-fit approximation to the experimental data, denoted $D_k^{\text{realized}}$. To determine which $D_k^{\text{realized}}$ is the best approximation to the true probability matrix $D$, we analyze the 
 \textit{training error} and \textit{testing error}. The training error for rank $k$ is simply the $\chi_k^{2 \ \text{train}}$ value obtained from Eq. (\ref{opt}) and it is used to determine how much a given model \textit{under-fits} the data. The testing error $\chi_k^{2 \ \text{test}}$ for a data set is given by
\begin{equation}
\label{chitest}
\chi_k^{2 \ \text{test}} = \sum_{i=1}^{m} \sum_{j=1}^{n} \left(\frac{F_{ij}^{\text{test}}-D_{k,ij}^\text{realized}}{\Delta F_{ij}^{\text{test}}}\right)^2,
\end{equation}
and is used to determine the \textit{predictive power} of a given model, which can be compromised by both under-fitting or \textit{over-fitting} the data. Here $F^{\text{test}}$ is the frequency matrix corresponding to the testing data set and $\Delta F_{ij}^{\text{test}}$ is the statistical uncertainty in $F_{ij}^{\text{test}}$. The rank-$k$ model that minimizes the testing error has the highest predictive power, and will be the model adopted as the best fit for our data.

\begin{figure}[h!]
     \centering
     \includegraphics[width=\columnwidth,keepaspectratio]{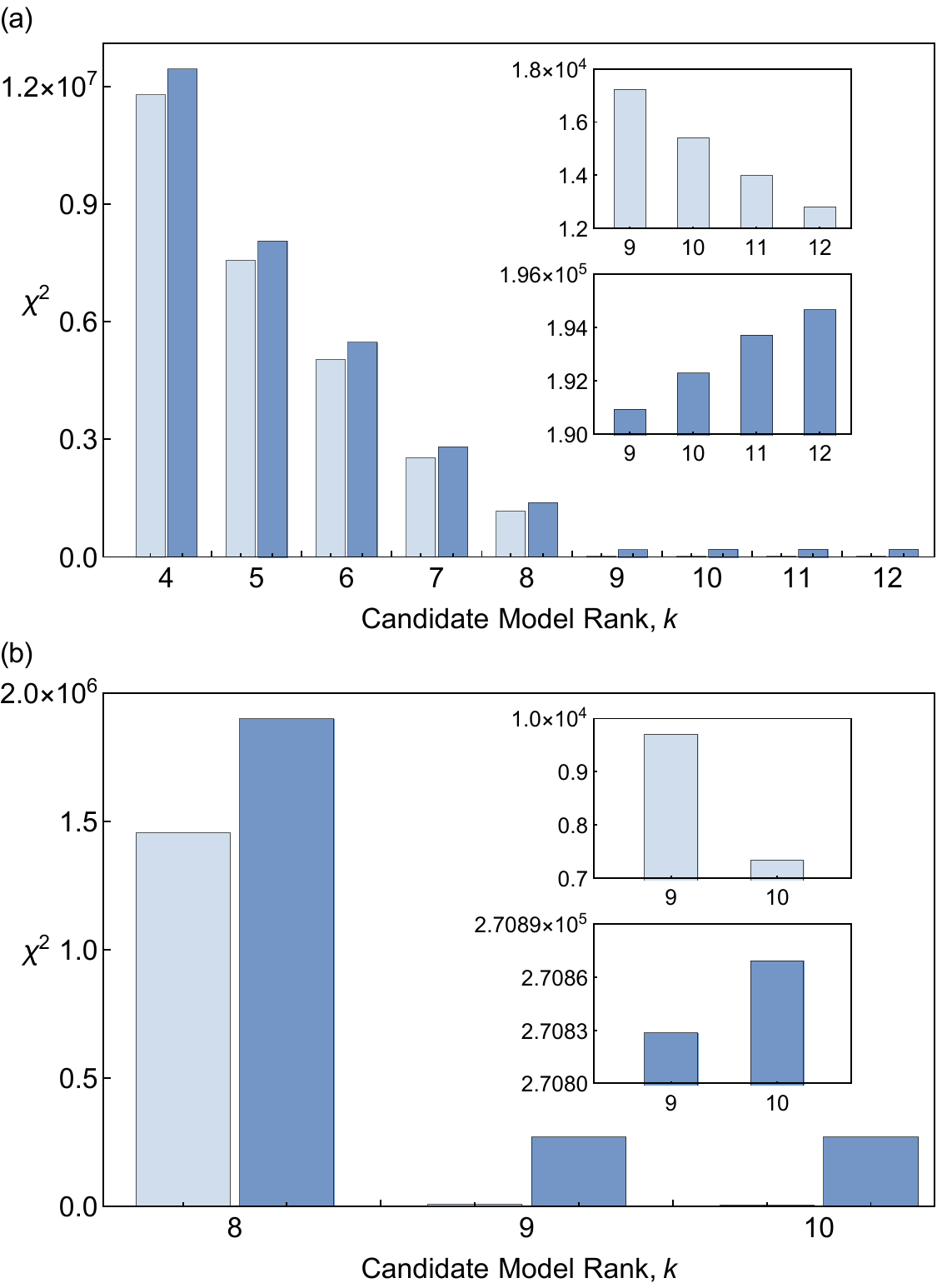}
     \caption{\edit{The training error (light blue) and testing error (dark blue) corresponding to various candidate model ranks $k$}. \textbf{(a) Training and testing errors for the $\mathbf{m = n = 100}$ 
      experiment}. Insets show training and testing error for model ranks 9 through 12. Models with rank $k<8$ severely under-fit the experimental data as indicated by the relatively high training error. Models with ranks $k>9$ begin to over-fit the experimental data, as indicated by the increase in the testing error \rob{accompanied by a decrease in the training error}. \textbf{(b) Training and testing errors for the $\mathbf{m = n = 415}$ 
     experiment.} The rank $k=8$ model under-fits the experimental data, while the rank $k=10$ model over-fits the data. For both experiments, we conclude that the best-fit GPT 
      is rank $k=9$ as it has the highest predictive power.}
     \label{fig:chiplots}
 \end{figure}

For the $100 \times 100$ Haar-distributed experiment, the training and testing errors for different ranks are shown in Fig. \ref{fig:chiplots}$(a)$. The training error (shown in blue) decreases as the candidate rank $k$ increases from 2 to 9.  It continues to decrease as the rank increases from 9 to 12, but at a slower rate.  The testing error (shown in red) also decreases as the candidate rank $k$ increases from 2 to 9, however, unlike the training error, it begins to rise again as the rank ranges from 9 to 12.  Since the rank-9 model is the one that is found to minimize the testing error, it is the one that is identified as having the maximum predictive power. Furthermore, the trends just described are precisely what one expects if the rank-9 model is the transition point between underfitting and overfitting, as we now explain.

Recall that the set of probability matrices that can be realized by models of rank $k-1$ is a {\em strict subset} of what can be realized by models of rank $k$ (the rank-$(k-1)$ models have strictly fewer parameters than the rank-$k$ models).  If the rank-$(k-1)$ model underfits the data relative to the rank-$k$ model, so that it has a higher training error and also a higher testing error, then every model with rank {\em less than} $k-1$ only further underfits the data relative to the rank-$k$ model and so necessarily has training and testing errors that are higher than or equal to those of the rank-$(k-1)$ model. Meanwhile, if the rank-$k$ model {\em overfits} the data relative to the rank-$(k-1)$ model (i.e., its greater parametric freedom causes it to fit to statistical fluctuations in the training data), so that it has a lower training error but a higher testing error, then every model with rank {\em greater than} $k$ only further overfits the data relative to the rank-$k$ model, implying that its training error can only fall further and its testing error \mike{tends to} rise higher.  The change in the rate of decrease of the training error after the point where the testing error is minimized is also expected.  Prior to that point, an increase in rank makes a big difference to the ability of the model to fit the data.  After that point, the additional parametric freedom can cause the model to fit to fluctuations in the training data, but this can only yield a slightly lower training error.  

In summary, the profile of the training and test errors that we observed is precisely what one expects under the assumption of the correctness of quantum theory: the rank $k=9$ model is the one at which there is a transition from underfitting to overfitting. The degree of underfitting monotonically increases as one decreases the rank below 9 and the degree of overfitting monotonically increases as one increases the rank above 9.

\begin{figure}[hbt!]
     \centering
     \includegraphics[width=\columnwidth,keepaspectratio]{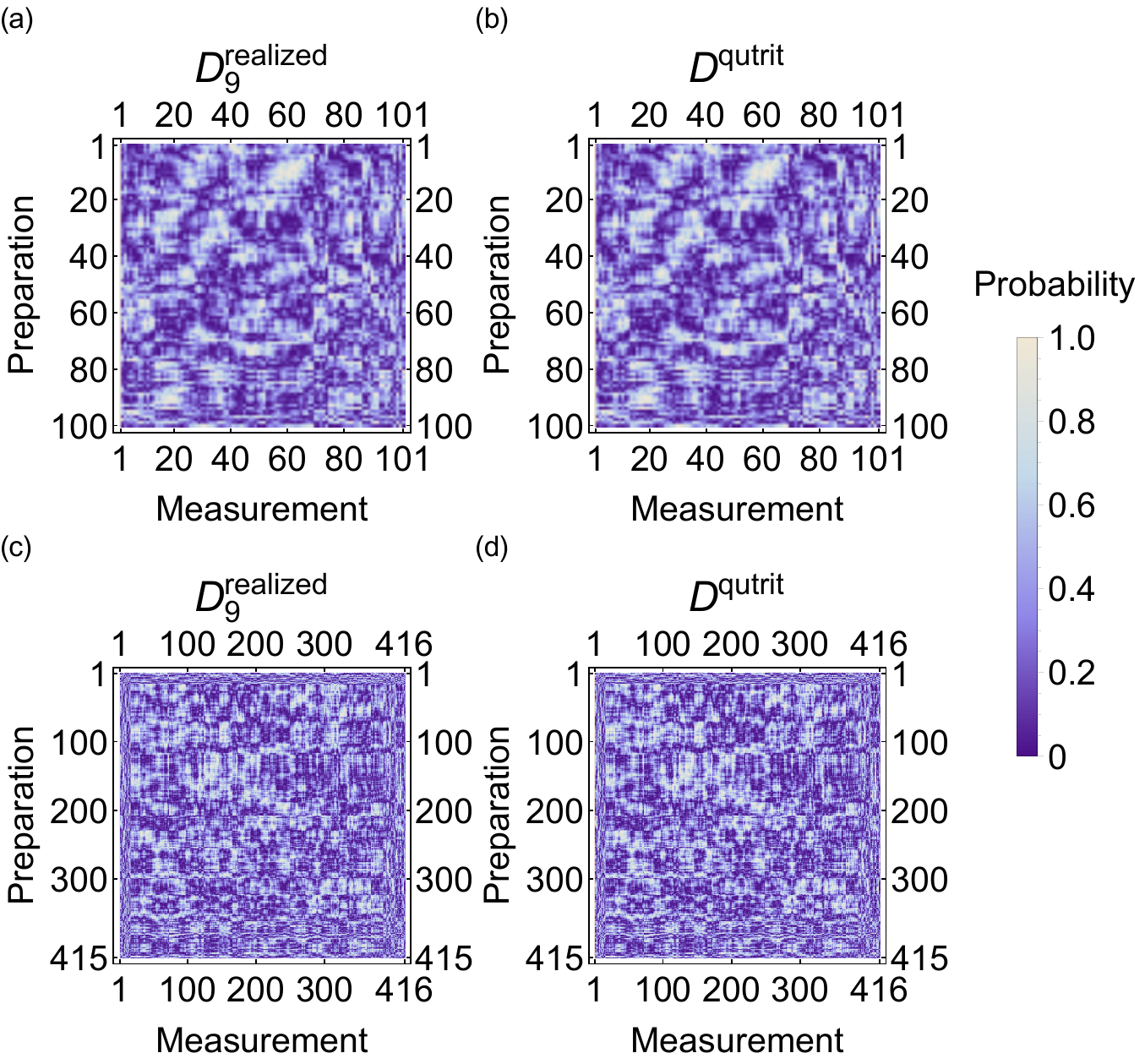}
     \caption{Comparison between the best-fit realized probability matrices $D_9^{\text{realized}}$ and the predicted quantum mechanical probability matrices $D^{\text{qutrit}}$ for our two experiments. (a) A matrix plot of $D_9^{\text{realized}}$ for the $m=n=100$ experiment. (b) A matrix plot of $D^{\text{qutrit}}$ corresponding to the preparation and measurement settings used in the $m=n=100$ experiment. The mean and standard deviation of the difference between $D_9^{\text{realized}}$ and $D_9^{\text{qutrit}}$ is -0.003 and 0.03, respectively. (c) A matrix plot of $D_9^{\text{realized}}$ for the $m=n=415$ 
      experiment. (d) A matrix plot of $D_9^{\text{qutrit}}$ for the $m=n=415$ 
       experiment. The mean and standard deviation of the difference between $D_9^{\text{realized}}$ and $D_9^{\text{qutrit}}$ is -0.006 and 0.03, respectively.}
     \label{fig:dmatrixcompare}
 \end{figure}

The best-fit probability matrix for our experimental data, therefore, is the rank-9 matrix $D_9^{\text{realized}}$. 
Figs.~\ref{fig:dmatrixcompare}(a)-(b) show a comparison between $D_9^{\text{realized}}$ and the theoretical probability matrix predicted by qutrit quantum mechanics, denoted $D^{\text{qutrit}}$. Fig.~\ref{fig:dmatrixcompare}(a) shows $D_9^{\text{realized}}$ as a matrix plot of probability values for the $m=n=100$ experiment. The random pattern of the matrix plot is a consequence of the Haar-distributed experimental configurations. Fig.~\ref{fig:dmatrixcompare}(b) shows a similar matrix plot for $D^{\text{qutrit}}$. The quantum mechanical probabilities were computed using $D_{ij}^{\text{qutrit}}=\text{Tr}[\rho_i Q_j]$ with $\rho_i$ and $Q_j$ being the same states and effects used to compute the wave plate angles for the experimental configurations. Figs.~\ref{fig:dmatrixcompare}(a)-(b) show strong agreement between the experimentally realized probability matrix and the probability matrix predicted by quantum theory with mean and standard deviation of the difference between $D_{ij}^{\text{realized}}$ and $D_{ij}^{\text{qutrit}}$ being -0.003 and 0.03, respectively. 

\subsection{Increasing the number of experimental configurations}
 
The $m=n=100$ experiment was useful for obtaining strong evidence that the dimension of the GPT underlying our experiment is 9, \rob{but the sampling of states and effects was not sufficiently dense to capture the geometries of the state and effect spaces}. In this section, we outline a method for increasing the number of GPT states and effects characterized in our experiment, so that we can obtain a clearer picture of the \rob{geometries.}

Simply increasing $m$ and $n$ \rob{and probing all state-measurement pairs} leads to a large increase in the number of experimental configurations, and consequently, an increase in the experimental run-time. However, to characterize a particular GPT state vector, one only needs to find its statistics on any tomographically complete subset of measurements. Similarly, to characterize any GPT effect vector, one only needs to implement it on a tomographically complete set of preparations. The cardinality of a tomographically complete set is the dimension $k$ of our GPT vector space. It follows that one should be able to characterize $m$ preparations and $n$ measurements with only $k(m+n)$ experimental settings, where $k$ is the dimension of the GPT. 

Despite the success of the first experiment in identifying the dimension of the GPT describing our data, it is prudent to not assume this result when conducting our second experiment. This means that although we expect we only need 9 states/effects to construct our tomographically complete sets, we use more than that to allow the new data to potentially refute the rank-9 conclusion of our first experiment. From the set of 24 normalized eigenvectors of the eight $3\times3$ Gell-Mann matrices (excluding identity), we choose the 15 that are distinct, denoted  $\{ |\psi_j\rangle\}_{j=1}^{15}$, as our tomographically overcomplete set of states. Our tomographically overcomplete set of measurements are simply the 15 binary-outcome measurements given by $\{Q_j, \bar{Q}_j\} = \{|\psi_j\rangle \langle \psi_j|, I-|\psi_j\rangle \langle \psi_i|\}$. We call the set of 15 tomographically overcomplete states (measurements) the \textit{fiducial} set and we use it to extend the number of states and effects we can characterize with our experiment. We sample 400 preparations and measurements according to the Haar distribution (in the same manner as our first experiment). We then pair each of the 15 fiducial states with the 400 Haar-distributed effects, and each of the 400 Haar-distributed states with the 15 fiducial effects. We also allow each fiducial state to be paired with each fiducial effect, adding $15^2$ state-effect pairings.  In total, the second experiment requires $15(400+400+15) = 12225$ state-effect pairings to be experimentally implemented, a modest $22\%$ increase relative to the 10000 pairings required in the first experiment, while allowing a characterization of 415 states and effects, which is more than four times the number of states and effects characterized in the first experiment. \rob{We again add the unit effect $\bm{u}$ by hand.}
 \rob{Our data is} arranged into a $415\times416$ frequency matrix $F$, where the bottom right $400\times400$ entries are unfilled. 

We run our second experiment twice back-to-back to obtain a training and test data set and perform a similar analysis to that of our first experiment. We choose to only analyze the candidate model ranks $k \in \{8,9,10\}$ \rob{because} our first experiment \rob{already indicated}
 that $k=9$ is likely to provide the best fit for our data and, as explained above, if $k=9$ is found to be the transition point between underfitting and overfitting in the range $k \in \{8,9,10\}$, then one can infer that ranks $k<8$ could only have exhibited more underfitting while ranks $k > 10$ could only have exhibited more overfitting. 

We find the best-fit matrix $D_k^{\text{realized}}$ for $F^{\text{train}}$ using Eq.~(\ref{opt}) and retrieve the training error for each model. For the $400 \times 400$ sub-matrix of unfilled entries in the frequency matrix, 
we take their weighting in the least-squares fit to be 0. Thus, the only constraint on the fit for the $400 \times 400$ unfilled entries is that each corresponding entry in $D_k^{\text{realized}}$ must be a valid probability. Once $D_k^{\text{realized}}$ is found for each candidate model, we compute the corresponding testing errors using Eq.~(\ref{chitest}). The training and testing errors for this data set were computed using only the elements of the frequency matrices that were actually measured in the experiment (i.e. the complement of the unfilled $400 \times 400$ sub-matrix of $F^{\text{train}}$ and $F^{\text{test}}$). 

The results of the training and testing error analysis for the fiducial experiment are shown in Fig.~\ref{fig:chiplots}(b). Comparing $k=8$ to $k=9$, the training and testing errors clearly indicate that the $k=8$ model under-fits the experimental data relative to the $k=9$ model, in agreement with our first experiment. Comparing $k=10$ to $k=9$, the fact that the training error decreases but the testing error increases indicates that the $k=10$ model is over-fitting the data relative to the $k=9$ model. We draw the conclusion that (in agreement with our first experiment) the model with the most predictive power for our data has rank $k=9$ and we adopt $D_9^{\text{realized}}$ as the best-fit probability matrix.

A comparison between $D_9^{\text{realized}}$ to the probability matrix predicted by qutrit quantum mechanics $D^{\text{qutrit}}$ for the fiducial experiment is shown in Fig. \ref{fig:dmatrixcompare}(c)-(d). Fig. \ref{fig:dmatrixcompare}(c) shows a matrix plot for the probabilities comprising $D_9^{\text{realized}}$ and Fig. \ref{fig:dmatrixcompare}(d) shows an analogous matrix plot for the quantum mechanically predicted probabilities. The narrow bands spanning the top and left sides of the matrix plots represent the preparations and measurements actually recorded in the experiment. These bands can be visually distinguished from the rest of the probabilities because the 15 fiducial states (effects) are more distinct from one another than the 400 states (effects) randomly sampled from the Haar distribution. The remainder of the probabilities in $D_9^{\text{realized}}$ are filled in from the weighted least squares fit. We again observe a strong agreement between the experimentally realized probability matrix and the probability matrix predicted by quantum theory with mean and standard deviation of the difference between $D_{ij}^{\text{realized}}$ and $D_{ij}^{\text{qutrit}}$ of -0.006 and 0.03, respectively.

\subsection{Constraining the shapes of the state and effect spaces}

We now discuss how to decompose the $m\times n$ best-fit probability matrix $D_9^{\text{realized}}$ into an $m\times k$ matrix $S^{\rm realized}$ of realized GPT state vectors and a $k\times n$ matrix $E^{\rm realized}$ of realized effect vectors. Recall that a decomposition $D = SE$ is not unique, because for any invertible matrix $\Lambda$, one also has $D=(S\Lambda) (\Lambda^{-1} E)$.  The choice of $\Lambda$ is merely a convention for how to represent the GPT state and effect spaces geometrically. We expect that because the dimension of our experimentally realized GPT is 9, we will be able to recover GPT states and effects that resemble the sets of quantum states and effects used to generate our experimental preparations and measurements. Thus, we seek to identify the choice of $\Lambda$ that yields the decomposition of $D_9^{\text{realized}}$ that best approximates the quantum states and effects used to generate our experimental wave plate angles (specifically, the generalized Bloch representations of these states and effects). Identifying this decomposition will allow us to more easily compare our realized GPT spaces to those predicted by quantum theory. 

 We begin by finding an initial valid decomposition of $D_9^{\text{realized}}$. To obtain this initial decomposition, we follow the method outlined in Appendix D of Ref.~\cite{Mazurek2017}. This method is summarized as follows: We first take the $QR$ decomposition of $D_9^{\text{realized}}$, where $Q$ is an $m \times m$ unitary matrix, and $R$ is an $m \times n$ upper-right triangular matrix. We then perform the singular value decomposition $QR = U\Sigma V^{T}$ and partition the result such that the $m \times k$ matrix of GPT states can be written as $S^{\prime} = U \sqrt{\Sigma}$ and the $k\times n$ matrix of GPT effects obtained from this decomposition can be written as $E^{\prime} = \sqrt{\Sigma} V^{T}$.

We define the matrices $S^{\text{qutrit}}$ and $E^{\text{qutrit}}$ to be those that consist of the Haar-distributed (and fiducial) qutrit state and effect vectors used to generate the wave plate angles for the second experiment. Given the initial decomposition $D_9^{\text{realized}} = S^{\prime}E^{\prime}$, the goal is to find an invertible $k\times k$ matrix $\Lambda$ such that $S^{\prime}\Lambda$ is as close as possible to $S^{\text{qutrit}}$. Therefore, we wish to find the matrix $\Lambda$ that minimizes the cost function
\begin{equation}
\label{trans}
\begin{aligned}
    \textrm{minimize} \quad & \chi^2 = \sum_i \sum_{\alpha} [ S_{i\alpha}^{\text{qutrit}}-(S^{'}\Lambda)_{i\alpha}]^2, \\
    \textrm{subject to} \quad & \Lambda \quad \text{invertible}
\end{aligned}
\end{equation}
where $i \in \{1,2,\dots, 415\}$ and \rob{$\alpha \in \{0,\dots, 8\}$}.

If $\Lambda_{\star}$ is the optimal transformation matrix computed from Eq.~(\ref{trans}), we take it as our conventional choice of $\Lambda$ in the decomposition $D_9^{\rm realized}= (S^{\prime}\Lambda) (\Lambda^{-1} E^{\prime})$ and thereby take
 $S^{\text{realized}} = S^{'}\Lambda_{\star}$ and $E^{\text{realized}} = \Lambda_{\star}^{-1}E^{'}$ to be the matrices defining our experimentally realized GPT state and effect vectors. This choice facilitates the project of looking for potential deviations from quantum theory.

The {\em realized} GPT state space, denoted $\bm{\mathcal{S}}^{\text{realized}}$, describes the states that were in fact 
\rob{realized} by the experiment.  
 The convex hull of the rows of the matrix $S^{\text{realized}}$ yields our best estimate of $\bm{\mathcal{S}}^{\text{realized}}$.  It is merely an estimate due to the fact that $S^{\text{realized}}$ is based on the relative frequencies in a finite run, rather than from long-run probabilities.
Similarly,  the {\em realized} GPT effect space, denoted $\bm{\mathcal{E}}^{\text{realized}}$, describes the effects that were in fact 
\rob{realized} by the experiment.  The procedure by which we obtain our best estimate of $\bm{\mathcal{E}}^{\text{realized}}$ is as follows. We begin by taking the set of GPT effect vectors defined by the columns of $E^{\text{realized}}$.  Next, we find the \rob{set of complements of these},
 i.e., the set of vectors ${\bf u} - {\bf e}$ for each ${\bf e}$ corresponding to a column of $E^{\text{realized}}$.  Finally, we identify the convex hull of these two sets.  Note that because the unit effect vector ${\bf u}$ is included as the first column of $E^{\text{realized}}$, it follows that the zero effect vector ${\bf 0}$ is included in the \rob{set of complements of effect vectors}, and so the unit and zero effect vectors are necessarily included in the convex hull.

The set of GPT effect vectors that are logically consistent with the realized GPT state vectors, termed the {\em consistent} effect space and denoted $\bm{\mathcal{E}}^{\text{consistent}}$, is the set of all vectors $\bm{e}$ such that $\forall \bm{s} \in \bm{\mathcal{S}}^{\text{realized}}: 0\leq \bm{s}\cdot \bm{e} \leq 1$. For simplicity, we write $\bm{\mathcal{E}}^{\text{consistent}} = \text{dual}(\bm{\mathcal{S}}^{\text{realized}})$. Analogously, the set of GPT state vectors that are logically consistent with the realized GPT effect vectors, termed the {\em consistent} state space and denoted $\bm{\mathcal{S}}^{\text{consistent}}$, is the set of all vectors $\bm{s}$ such that $\forall \bm{e} \in \bm{\mathcal{E}}^{\text{realized}}: 0\leq \bm{s}\cdot \bm{e} \leq 1$ and $\bm{s}\cdot \bm{u} = 1$. We write $\bm{\mathcal{S}}^{\text{consistent}} = \text{dual}(\bm{\mathcal{E}}^{\text{realized}})$. 

We now explain how the realized and consistent state (effect) space obtained from our experiment relates to the true state (effect) space governing nature. 
First, the state and effects that are realized experimentally necessarily involve additional noise relative to what is  possible in the true GPT.  As such, we expect all of the states and effects we realize to be slightly noisy versions of the ones that are stipulated to be possible by the true GPT.  Geometrically, this means that the realized state and effect vectors should be slightly {\em contracted}, in directions orthogonal to the normalization axis, relative to their noiseless counterparts. And this in turn implies that the state and effect vectors that are {\em logically consistent} with the ones we realized should be slightly {\em elongated} in these directions relative to their noiseless counterparts. 
Second, the fact that one can only realize a {\em finite} number of states and effects in the experiment implies that the realized state and effect spaces will be convex polytopes even if the true state and effect spaces are not.  The same is true of the logically consistent spaces that one obtains from the realized spaces. The extent to which the realized and consistent spaces approximate the true and logical spaces depends on how densely sampled the full set of states and effects is. 

The experimental limitations described above imply that the realized spaces are expected to be {\em strictly contained} within the true spaces, $\bm{\mathcal{S}}^{\text{realized}} \subset \bm{\mathcal{S}}$ and $\bm{\mathcal{E}}^{\text{realized}} \subset \bm{\mathcal{E}}$, while the consistent spaces strictly contain the logical spaces, $\bm{\mathcal{S}}^{\text{logical}} \subset \bm{\mathcal{S}}^{\text{consistent}}$ and $\bm{\mathcal{E}}^{\text{logical}} \subset \bm{\mathcal{E}}^{\text{consistent}}$. It follows that the true GPT state (effect) space governing our three-level system must lie somewhere between the realized and consistent GPT state (effect) space inferred from our experiment. More precisely, $\bm{\mathcal{S}}$ must satisfy $\bm{\mathcal{S}}^{\text{realized}} \subseteq \bm{\mathcal{S}} \subseteq \bm{\mathcal{S}}^{\text{consistent}}$, which is equivalent to the constraint that $\bm{\mathcal{E}}$ must be such that $\bm{\mathcal{E}}^{\text{realized}} \subseteq \bm{\mathcal{E}} \subseteq \bm{\mathcal{E}}^{\text{consistent}}$.  (Recall also that $\bm{\mathcal{S}}$ and $\bm{\mathcal{E}}$ are not independently specifiable. For instance, once $\bm{\mathcal{S}}$ is specified, the possibilities for $\bm{\mathcal{E}}$ are constrained to satisfy $\bm{\mathcal{E}} \subseteq {\rm dual}(\bm{\mathcal{S}})$.) Hence, one does not {\em uniquely} identify the true state and effect spaces; one only delimits the scope of possibilities for what they might be.

 Finally, to plot 3-dimensional projections of $\bm{\mathcal{S}}^{\text{realized}}$ and $\bm{\mathcal{E}}^{\text{realized}}$, we rely on the fact that these convex sets are polytopes. We first identify the vectors describing the vertices of these polytopes. We then compute the projections of each of these vertices. Finally, we plot the convex hull of the resulting sub-vectors corresponding to each projection. 
 The definitions of $\bm{\mathcal{S}}^{\text{consistent}}$ and $\bm{\mathcal{E}}^{\text{consistent}}$ provided above are known as the \textit{inequality representations}. However, in order to plot projections of $\bm{\mathcal{S}}^{\text{consistent}}$ and $\bm{\mathcal{E}}^{\text{consistent}}$ in a similar manner to how we plot projections of $\bm{\mathcal{S}}^{\text{realized}}$ and $\bm{\mathcal{E}}^{\text{realized}}$, we require the so-called \textit{vertex representations} of the spaces, which lists all the vertices that specify their convex hulls. Converting from an inequality representation to a vertex representation is highly non-trivial, and requires solving the \textit{vertex enumeration problem}. Fortunately, the solution to the problem can be computed using an algorithm first developed by Avis and Fukada~\cite{Avis1991}. Calculation of the vertex representations of our consistent spaces was performed using the lrs package in C that implements a modified version of the vertex enumeration algorithm~\cite{lrs}.

\robnew{We end this section with some comments on how experimental imperfections manifest themselves in this data analysis technique. 

We begin by considering this question under the assumption of the correctness of quantum theory.  Although an experimentalist may aim to achieve the ideal of {\em pure} states and {\em projective} measurements, what is realized experimentally always deviates from this ideal.  It is a critical feature of bootstrap GPT tomography that the rows of the matrix $S^{\text{realized}}$ and the columns of the matrix  $E^{\text{realized}}$ (which ultimately determine $\bm{\mathcal{S}}^{\text{realized}}$ and $\bm{\mathcal{E}}^{\text{realized}}$) are characterizations of the preparations and measurements that were {\em actually realized} in the experiment, and not characterizations of the idealized preparations and measurements that the experimentalist was targeting.  

It is  useful to consider two distinct ways in which a realized procedure might deviate from the ideal that was being targeted: it might differ in a systematic and consistent way from the target (i.e., inaccuracy) or it might exhibit statistical fluctuations away from the target (i.e. imprecision).
In our experiment, an example of {\em inaccuracy} is if the wave-plate angles that are implemented for a given preparation procedure differ in a systematic and consistent way (across all repetitions of the procedure) from those that would achieve the particular pure state being targeted.  In this case, our tomography scheme would return a best-fit GPT state that is 
rotated in Bloch space (i.e., the qutrit analogue of the Bloch ball) relative to the target.  An example of {\em imprecision} is if, in the ensemble of repetitions of a given preparation procedure, a wave-plate angle exhibits statistical fluctuations away from the targeted value.  In this case, our tomography scheme would return a best-fit GPT state 
that was a noisy version of the target, 
for instance, one that is {\em contracted} (in Bloch space)
relative to the pure state being targeted.  

If one is willing to make strong assumptions about the nature of the imperfections in one's experiment, then one might hope to process the raw data to compensate for these and thereby achieve a characterization of the ideal procedures that one was targeting. (Background subtraction is an example of such a processing.)  Bootstrap GPT tomography, however, simply analyzes the raw data, rather than a processed version thereof, and allows one to {\em infer} the nature of the imperfections.




Making such inferences about experimental imperfections, however, is only possible relative to some assumption about what the correct operational theory is.
This is because it is only under such an assumption that an experimentalist can have {\em any expectation} about what GPT states and effects will characterize some experimental procedure (even an idealized one).  For example, under the assumption of the correctness of {\em quantum theory}, one expects certain idealized preparation procedures to be on the surface of the state space dictated by quantum theory (i.e. Bloch space), such that rotations and contractions of these GPT states relative to one’s expectations can be used to infer inaccuracies and imprecisions in the actually realized procedures relative to their idealized counterparts.   

Given that the purpose of this article is to find a way to analyze the experimental data {\em without} making assumptions about which theory is correct, it is imperative to {\em not} make any assumptions about what GPT states describe the experimental procedures one is implementing (or putative idealized versions thereof).  It follows that the data analysis scheme one uses {\em must} be such that it returns a characterization of the actually realized procedures, as bootstrap GPT tomography does.
To put it another way: in a theory-agnostic tomography scheme, it is not possible to imagine a processing of the raw data that can compensate for imperfections relative to some putative ideal because one is not allowed to make assumptions about the nature of this ideal. 
}

\section{Results}

\begin{figure*}[hbt!]
     \centering
     \includegraphics[width=\textwidth,keepaspectratio]{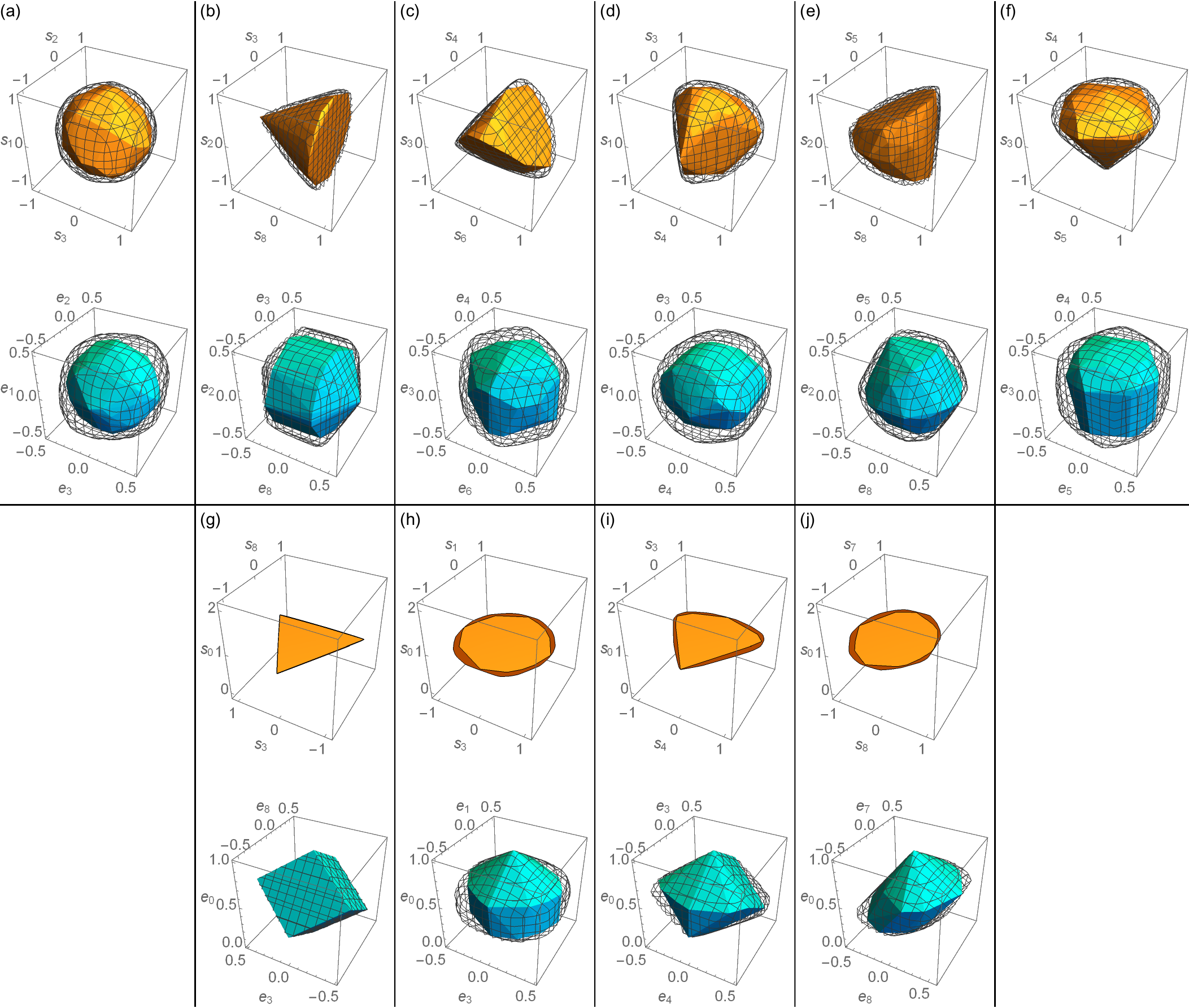}
     \caption{\edit{A sampling of 3-dimensional projections of the realized GPT state space $\bm{\mathcal{S}}^{\text{realized}}$ (top polytope of each subfigure, displayed in yellow) and of the realized GPT effect space $\bm{\mathcal{E}}^{\text{realized}}$ (bottom polytope of each subfigure, displayed in cyan)}. The mesh polytopes surrounding each of the yellow polytopes depicts 3-dimensional projections of the consistent GPT state space $\bm{\mathcal{S}}^{\text{consistent}}$, while the mesh polytopes surrounding each of the cyan polytopes depicts 3-dimensional projections of the consistent GPT effect space $\bm{\mathcal{E}}^{\text{consistent}}$.  The true state and effect spaces of the GPT describing nature must lie somewhere between the realized and consistent GPT spaces. The specific projections depicted in (a)-(i) are the same as in Fig.~\ref{fig:maxextentstate}.}
     \label{fig:expmaxextenteffect}
 \end{figure*}

In Fig.~\ref{fig:expmaxextenteffect}, we present various 3-dimensional projections of the realized and consistent spaces we found in our experiment. Specifically, we present the same projections that are showcased in Fig.~\ref{fig:maxextentstate} for ease of comparison with the latter. For a given 3-dimensional subspace, we plot the projection of $S^{\text{realized}}$ into this subspace as a yellow polytope, and the projection of $E^{\text{realized}}$ as a cyan polytope. The corresponding projections of the spaces $\bm{\mathcal{S}}^{\text{consistent}}$ and $\bm{\mathcal{E}}^{\text{consistent}}$ are plotted as mesh polytopes alongside the projections of $\bm{\mathcal{S}}^{\text{realized}}$ and $\bm{\mathcal{E}}^{\text{realized}}$. For the state space projections that include the normalization component, depicted in Fig.~\ref{fig:expmaxextenteffect}(g)-(j), the consistent state spaces are plotted in dark orange rather than as a mesh. 


A visual comparison of Fig.~\ref{fig:expmaxextenteffect} to Fig.~\ref{fig:maxextentstate} reveals that our results do conform broadly with the hypothesis that the qutrit state (effect) space lies between the realized and consistent state (effect) spaces obtained in our experiment. Strictly speaking, consistency holds if some {\em invertible linear transformation} of the qutrit space lies between the realized and consistent space. We shall return to the question of consistency with quantum theory at the end of this section.

The question of whether the no-restriction hypothesis holds for the true GPT governing nature is the question of whether all logically possible states (effects) are included in the state (effect) space.  In other words, it is the question of whether there is a gap between $\bm{\mathcal{S}}$ and $\bm{\mathcal{S}}^{\rm logical}$ or between $\bm{\mathcal{E}}$ and $\bm{\mathcal{E}}^{\rm logical}$. As noted above, however,  experimental noise and finite sampling imply that the gap between the realized and consistent spaces is expected to be strictly larger than the gap between the true and logical spaces.  Seeing a nontrivial gap of the former type, therefore, does not imply that a gap of the latter type exists.  Nonetheless, one can put an {\em upper bound} on possible violations of the no-restriction hypothesis.

The gap between  $\bm{\mathcal{S}}^{\text{realized}}$ and $\bm{\mathcal{S}}^{\text{consistent}}$ \rob{therefore} delimits the scope of possible deviations from quantum theory. In particular, they bound how far any candidate GPT describing our three-level system may deviate from the no-restriction hypothesis. 
 We will here focus on bounding the extent to which the no-restriction hypothesis might fail based on the gap between $\bm{\mathcal{S}}^{\text{realized}}$ and $\bm{\mathcal{S}}^{\text{consistent}}$.  We turn now to the question of how to quantify this gap. 

In the investigation of two-level systems undertaken in Ref.~\cite{Mazurek2017}, the gap was quantified by the ratio of the {\em volumes} of $\bm{\mathcal{S}}^{\text{realized}}$ and $\bm{\mathcal{S}}^{\text{consistent}}$.  This approach is not feasible here---the fact that the state spaces are 9-dimensional and the fact that the consistent state space has a very large number of vertices made it \mike{intractable} to compute the volume ratio in a straightforward manner. We have opted, therefore, to consider instead a ratio of {\em linear dimensions}, specifically, the distances from \rob{the `centre' of the space of normalized states, the vector $(1,0,\dots,0)$, }
 to the surface of the realized and consistent state spaces along some direction. 
 As this ratio of distances will vary with the direction being considered, we quantify the gap by averaging the ratio over a large sampling of random directions.
We consider directions defined by the generalized Bloch vectors of pure quantum states,
and we sample these according to the Haar measure.
 For each ray, we divide the norm of the vector extending to the intersection point with  $\bm{\mathcal{S}}^{\text{realized}}$ by the norm of the vector extending to the intersection point with $\bm{\mathcal{S}}^{\text{consistent}}$ to obtain our \textit{linear dimension ratio}. For a set of 1000 rays, the average linear dimension ratio between  $\bm{\mathcal{S}}^{\text{realized}}$ and $\bm{\mathcal{S}}^{\text{consistent}}$ is computed to be $0.80$, with an associated standard deviation of $0.04$.  Any candidate GPT describing our three-level system that admits an average linear dimension ratio much smaller than $0.80$ \rob{exhibits a greater failure of}
  the no-restriction hypothesis \rob{than is}
 consistent with our experimental findings.
 
 Note that the average linear dimension ratio within each of the various {\em 3-dimensional projections} of the realized and consistent spaces is not related in a simple manner to the average linear dimension ratio for the full-dimensional spaces.  This is because projections need not correspond to sections that pass through the origin. Nonetheless, a visual inspection of Fig.~\ref{fig:expmaxextenteffect} reveals that the average linear dimension ratio for many of these projections might well be close to 0.80.

To conclude this section, we return to the question of whether an invertible linear transformation of the qutrit state space $\bm{\mathcal{S}}^{\text{qutrit}}$ fits between the realized and consistent state spaces, $\bm{\mathcal{S}}^{\text{realized}}$ and $\bm{\mathcal{S}}^{\text{consistent}}$. The question turns out to be more nuanced than it appears at first glance.  What complicates matters is that the rows of the matrix $S^{\rm realized}$ are, strictly speaking, only finite-sample {\em estimates} of the states in $\bm{\mathcal{S}}^{\rm realized}$.  Recall that, by definition, a given state vector in $\bm{\mathcal{S}}^{\rm realized}$ specifies the long-run probabilities that would be obtained by the associated realized preparation procedure, and finite-run relative frequencies need not coincide with these long-run probabilities. Finite-run estimates of the realized state vectors will, therefore, 
\rob{exhibit fluctuations away} from the realized state vectors. In particular, the finite-run estimates of certain realized state vectors might be anomalously long.
\rob{Similarly, the finite-run estimates of certain realized {\em effect} vectors might be anomalously long, implying} that the finite-run estimates of certain {\em consistent} state vectors might be anomalously {\em short}. 
 Together these two anomalies can lead to a situation wherein \robn{$\bm{\mathcal{S}}^{\text{qutrit}}$ does not in fact fit between the estimate of $\bm{\mathcal{S}}^{\text{realized}}$ and the estimate of $\bm{\mathcal{S}}^{\text{consistent}}$.}


\robn{Given that the geometry of the qutrit state space is nontrivial, and given that our estimate of the consistent state space has a very large number of vertices, it is not straightforward to solve this decision problem. We therefore resort to a heuristic assessment of the question. 
We again consider the 1000 rays sampled according to the Haar measure from the directions defined by the generalized Bloch vectors of pure qutrit quantum states.  
Let $l$ be an index that runs over these rays. For each $l$, define  $\bm{s}_{l}^{\text{realized}}$ to be the vector in the direction of the $l$th ray that lies on the boundary of
$\bm{\mathcal{S}}^{\text{realized}}$, and define  $\bm{s}_{l}^{\text{consistent}}$ to be the one  that lies on the boundary of $\bm{\mathcal{S}}^{\text{consistent}}$.
Because all pure qutrit states have generalized Bloch vectors of the same norm, if the longest of the $\bm{s}_{l}^{\text{realized}}$ is shorter than the shortest of the $\bm{s}_{l}^{\text{consistent}}$, i.e., if $\max_l ||\bm{s}_{l}^{\text{realized}}||\le \min_l ||\bm{s}_{l}^{\text{consistent}}||$, then there is a linear transformation of $\bm{\mathcal{S}}^{\text{qutrit}}$ (in particular, one that preserves the relative norm of pure states) which is straddled by  our estimate of $\bm{\mathcal{S}}^{\text{realized}}$ and our estimate of $\bm{\mathcal{S}}^{\text{consistent}}$ at least along the 1000 random directions.  We find $\max_l ||\bm{s}_{l}^{\text{realized}}|| = 1.11$ and $\min_l ||\bm{s}_{l}^{\text{consistent}}|| = 1.13$, so the straddling condition does hold for the 1000 random directions. 
In summary, our experimental results seem to be consistent with the hypothesis that quantum theory is the true GPT underlying our experiment. }

\section{Conclusions}

In this paper, we reconstructed the GPT state and effect spaces for a three-level photonic system directly from experimental data. The tomographic scheme utilized is \robnew{bootstrap} in that it does not require any prior characterizations of the preparations and measurements used in the experiment. Furthermore, neither our experimental scheme nor our analysis assumed the correctness of quantum theory. From the statistics obtained in our experiment, we were able to infer that \rob{if our sets of states and effects are tomographically complete then the} 
 dimension of the underlying GPT for our system is 9. This \rob{inference} was accomplished by computing the training and testing error (Eqs.~(\ref{opt}) and (\ref{chitest})) for the best-fit probability matrices of different ranks, \rob{and identifying the model that tests best}. Although our experiment provided an opportunity to discover that the dimension of the GPT governing our three-level system differed from the quantum mechanical predictions, our analysis did not yield any evidence in favour of such a deviation.

We have also identified the scope of possible shapes of the true GPT state and effect spaces, assuming they are indeed 9-dimensional. The true GPT state space must lie between the realized GPT state  space $\bm{\mathcal{S}}^{\text{realized}}$  and the consistent GPT state space $\bm{\mathcal{S}}^{\text{consistent}}$. Similarly, the true GPT effect space must lie between the realized GPT effect space $\bm{\mathcal{E}}^{\text{realized}}$ and the consistent GPT effect space $\bm{\mathcal{E}}^{\text{consistent}}$.
We displayed the geometries of our \rob{estimates of the} realized and consistent state and effect spaces as 3-dimensional projections of the full 9-dimensional spaces in Fig.~\ref{fig:expmaxextenteffect}. In addition, we were able to place quantitative bounds on how much any proposed alternative to quantum theory may violate the no-restriction hypothesis by identifying the average \rob{ratio of linear dimensions of $\bm{\mathcal{S}}^{\text{realized}}$ and of $\bm{\mathcal{S}}^{\text{consistent}}$}
 to be $0.80$. Even though our experimental data provides some room for violations of the no-restriction hypothesis, it is notable that we observe good agreement in the shapes of our experimental GPT spaces when comparing them to the quantum mechanical predictions. 
 
The results obtained in this paper provide constraints on the scope of possible deviations from quantum theory. There are many avenues to explore in conducting further experiments of this kind. The obvious extension of this work would be to undertake a GPT characterization of higher-dimensional systems, or of  composite systems. For example, one might undertake a GPT characterization of a pair of qubits and test the validity of principles that concern such systems, such as the principle of local tomography. Another interesting possible extension of this work is to undertake a GPT characterization of channels by performing the GPT equivalent of quantum process tomography. 
    
The results of our analysis and that of Ref.~\cite{Mazurek2017} demonstrate the utility of the GPT framework in adjudicating between quantum theory and its alternatives in the landscape of possible physical theories.

\section*{Acknowledgements}
The authors would like to thank Tim Hill for outlining some theoretical simulations relating to this experiment and Patrick Daley for assisting with the experimental setup and for useful discussions regarding model selection techniques. The authors would also like to thank Elie Wolfe for assisting with the linear dimension ratio analysis. Finally the authors thank Sandra Cheng, Sacha Schwarz, Aldo Pasos, and Satchel Armena for fruitful discussions. This research was supported in part by the Natural Sciences and Engineering Research Council of Canada (NSERC), the Ontario Graduate Scholarship (OGS), Industry Canada, the Canadian Foundation for Innovation (CFI), the Canada First Research Excellence Fund (CFREF), and the Perimeter Institute for Theoretical Physics. Research at Perimeter Institute is supported in part by the Government of Canada through the Department of Innovation, Science and Economic Development Canada and by the Province
of Ontario through the Ministry of Colleges and Universities.

\bibliography{GPTpaper}

\end{document}